\documentclass[
preprint,
showpacs,amsmath,amssymb,aps,pra,nofootinbib]{revtex4-1}

\usepackage{graphicx}
\usepackage{amsmath}

\usepackage[dvipsnames]{xcolor}
\usepackage{color}
\usepackage[pdfpagelabels,pdftex,bookmarks,breaklinks]{hyperref}
\definecolor{darkblue}{RGB}{0,0,127} 
\definecolor{darkgreen}{RGB}{0,150,0}
\hypersetup{colorlinks, linkcolor=blue, citecolor=magenta, filecolor=red, urlcolor=blue}
\hypersetup{pdfpagemode=UseNone} 
\hypersetup{pdfstartview=FitH} 

\newcommand{\eref}[1]{(\ref{#1})}

\newcommand{\vac}{|{\rm vac.}\rangle}

\begin{document}

\title{Encoding qubits into harmonic-oscillator modes via quantum walks in phase space}

\author{Chai-Yu Lin, Wang-Chang Su, and Shin-Tza Wu}
\email{phystw@gmail.com} \affiliation{ Department of Physics,
National Chung Cheng University, Chiayi 621, Taiwan}

\date{\today}

\begin{abstract}
We provide a theoretical framework for encoding arbitrary logical
states of a quantum bit (qubit) into a continuous-variable quantum
mode through quantum walks. Starting with a squeezed-vacuum state of
the quantum mode, we show that quantum walks of the state in phase
space can generate output states that are variants of codeword
states originally put forward by Gottesman, Kitaev, and Preskill
(GKP) [Phys. Rev. A {\bf 64}, 012310 (2001)]. In particular, with a
coin-toss transformation that projects the quantum coin onto the
diagonal coin-state, we show that the resulting {\em dissipative}
quantum walks can generate qubit encoding akin to the prototypical
GKP encoding. We analyze the performance of these codewords for
error corrections and find that even without optimization our
codewords outperform the GKP ones by a narrow margin. Using the
circuit representation, we provide a general architecture for the
implementation of this encoding scheme and discuss its possible
realization through circuit quantum-electrodynamics systems.
\end{abstract}

\maketitle

\section{\label{sec:int}Introduction}
Computing based on quantum mechanical principles (i.e., quantum
computing) requires exquisite control of quantum systems
\cite{NC00}. Thanks to advancements in experimental techniques,
tremendous progress has been made for achieving this goal during the
past few years \cite{Ca17}. For large scale quantum computing, it is
indispensable to have an architecture that enables efficient
detection and correction of errors during the computing \cite{Li13}.
Recently, there has been significant progress towards this direction
in the field of continuous-variable (CV) measurement-based quantum
computing, which seeks to achieve quantum computing by sequence of
adaptive local measurements over highly entangled resource states in
a state space with continuous spectrum \cite{Zh06,Me06}. In
particular, Menicucci has shown that fault-tolerant quantum
computing can be achieved in this scheme provided resource states
with squeezing above $20.5$ dB are available \cite{Me14}. Recently,
with the aid of topological codes, this squeezing threshold has been
reduced to less than $10$ dB \cite{Fu17,Fu18}. Essential to these
breakthroughs is a quantum error-correcting scheme due to Gottesman,
Kitaev, and Preskill (GKP) \cite{GKP01}. In this approach, quantum
information is encoded through a ``hybrid" quantum bit (qubit)
embedded in the (quantum mechanical) phase space of a quantum
harmonic oscillator. Despite the importance of the GKP scheme,
existing proposals for the experimental generation of GKP qubits
remain to pose major challenges
\cite{Tr02,Pi04,Pi06a,Pi06b,Va10,Br13,Te16,Mo17,We18} (see, however,
the recent report in Ref.~\cite{Fl19}). In this paper we propose a
new scheme for preparing GKP qubits through quantum walks (QWs) of
an oscillator mode in phase space \cite{Ke03,Ma14}. As we will show,
our encoding scheme indeed also provides a framework for
experimentally accessing other general ``grid states", namely,
codeword states with grid-like structures in phase space.

In the CV approach to quantum computing, quantum information are
carried by quantum modes (aka ``qumodes") with the logical states
encoded via eigenstates of the canonical coordinates of the field
mode, which are usually likened to the position and momentum of a
harmonic oscillator \cite{BP03,BrvL05}. Decoherence of the qumode
then manifests as shift errors in these basis states. In order to
correct such errors, GKP propose to invoke ``hybrid" qubits that
consist of superposition of uniformly spaced position eigenstates
separated by $2\sqrt{\pi}$ \cite{GKP01}
\begin{eqnarray}
|0\rangle_L &=&\!\!\sum_{s=-\infty}^\infty\!\!|2s\sqrt{\pi}\,\rangle_x
= \frac{1}{\sqrt{2}}\!\sum_{s=-\infty}^\infty\!\!|s\sqrt{\pi}\,\rangle_p \, ,
\nonumber \\
|1\rangle_L &=& \!\!\sum_{s=-\infty}^\infty\!\!|(2s+1)\sqrt{\pi}\,\rangle_x
= \frac{1}{\sqrt{2}}\!\sum_{s=-\infty}^\infty\!\!(-1)^s |s\sqrt{\pi}\,\rangle_p ,
\label{ideal_GKP}
\end{eqnarray}
where $|x\rangle_x$ and $|p\rangle_p$ are, respectively, position
and momentum eigenstates. Thus the position-space wavefunctions for
the codewords $|0\rangle_L$ and $|1\rangle_L$ comprise combs of
delta functions located at, respectively, even and odd multiples of
$\sqrt{\pi}$. In the presence of shift errors, it is then possible
to correct sufficiently small errors in the encoded qubits through
position and momentum measurements \cite{GKP01}. However, the GKP
codeword states \eref{ideal_GKP} require infinite squeezing, and
hence infinite energy. In practice, therefore, one must approximate
\eref{ideal_GKP} with finitely squeezed states, such as uniformly
spaced Gaussian spikes modulated by Gaussian envelopes \cite{GKP01}
\begin{eqnarray}
{}_x\langle x|\tilde{l}\rangle_L &\propto&
\sideset{}{'}\sum_n \,e^{-\frac{n^2\pi\Delta_p^2}{2}} \,
\exp\!\left[-\frac{(x-n\sqrt{\pi})^2}{2\Delta_x^2}\right] ,
\nonumber\\
{}_p\langle p|\tilde{l}\rangle_L &\propto&
\sum_n (-1)^{nl} \,e^{-\frac{\Delta_x^2\,p^2}{2}} \,
\exp\!\left[-\frac{(p-n\sqrt{\pi})^2}{2\Delta_p^2}\right] ,
\label{GKP_wfs}
\end{eqnarray}
where $l=\{0,1\}$ are the logical bit values, $\sum_n'$ in the
position-space wavefunctions indicate summations over even/odd
integers $n$ for $l=0/1$, and $\Delta_x$, $\Delta_p$ specify the
widths of the Gaussians. It is our goal in the present paper to
provide experimentally feasible schemes for engineering approximate
GKP codewords such as \eref{GKP_wfs} by implementing QWs in phase
space for a qumode. In contrast to its classical counterpart, QW
takes place in accordance with a quantum coin that admits
superposition of orthogonal coin-states \cite{Ke03,Ma14}. Utilizing
QWs of a qumode initially in a squeezed vacuum state, we will
demonstrate that GKP-type codewords can be generated under
appropriate ``coin-toss rules". As we will show, by changing the
nature of the coin toss, one can attain GKP-type encodings with
different characteristics. Our approach thus offers not only a
promising pathway to the preparation of GKP qubits, but also opens
up new dimensions to the GKP encoding.

In the following, we will start in Sec.~\ref{sec:form} by first
explaining how the features of QWs in phase space for a qumode can
be exploited to encode a squeezed vacuum state into a GKP qubit. We
will then illustrate with two instances: One involving generic
unitary QWs, and the other {\em dissipative}, non-unitary QWs.
Performance of our codewords for error correction will then be
analyzed for the dissipative case. We will then discuss in
Sec.~\ref{sec:implm} the implementation for our encoding scheme by
first establishing a protocol using the circuit model and next
touching on its possible experimental realizations through circuit
quantum-electrodynamics systems. Finally, we conclude in
Sec.~\ref{sec:concl} with a summary and brief discussions for our
findings. For presentational clarity, we relegate technical details
and elaborate formulas to the Appendices.

\section{\label{sec:form} From quantum walk to the GKP encoding}
Let us consider one-dimensional (1D) QW in phase space for a qumode
with mode operator $\hat{a} = (\hat{x} + i\hat{p})/\sqrt{2}$, where
$\hat{x}$ and $\hat{p}$ are, respectively, the ``position" and
``momentum" quadrature operators of the qumode. From the commutation
relation $[\hat{a},\hat{a}^\dagger]=1$ for the mode operator, it
follows that $[\hat{x},\hat{p}]= i$, which corresponds to setting
$\hbar=1$ for us. For the QW, we will be concerned with
position-squeezed states for the qumode, which will be denoted as
$|q\rangle_r$. Here the subscript $r$ indicates the squeezing
parameter, $q$ is the expectation value $\langle\hat{x}\rangle$ of
the state in units of $\sqrt{2}$ times the QW step length
$\xi_d$.\footnote{\label{fn:sqrt2}Here the extra factor $\sqrt{2}$
comes from defining the Hermitian part of the mode operator
$\hat{a}$ to be $\hat{x}/\sqrt{2}$ (and hence $\hbar=1$). If one
uses instead $\hat{x}$ for the Hermitian part of $\hat{a}$ (thus
$\hbar=1/2$), such factor of $\sqrt{2}$ would then disappear. Here
we are following conventions used commonly in the GKP literatures.}
Each step of the QW is conditioned on the configuration
$|\epsilon\rangle$ of a two-state quantum coin with
$\epsilon=\{R,L\}$ corresponding to rightward ($R$) and leftward
($L$) displacements by one single step length. More precisely, in
terms of the phase-space displacement operator $\hat{D}(\xi)\equiv
\exp\{\xi(\hat{a}^\dagger-\hat{a})\}$ for real $\xi$ and the
squeezing operator $\hat{S}(r)\equiv\exp\{\frac{r}{2}
(\hat{a}^2-\hat{a}^{\dagger 2})\}$ for real $r$, we will be
considering squeezed coherent states
\begin{eqnarray}
|q\rangle_r
\equiv \hat{D}(q\,\xi_d)\, \hat{S}(r) \vac
\label{qx}
\end{eqnarray}
with $|{\rm vac.}\rangle$ the vacuum state of the qumode. Therefore,
in the language of QW, the product state
$|q\rangle_r|\epsilon\rangle$ indicates a walker at position
$q\times\sqrt{2}\xi_d\equiv q\,x_d$ with a coin configuration
$\epsilon$ (see footnote \ref{fn:sqrt2}).

Let us suppose the QW has the coin-toss operator $\hat{\cal C}$. The
corresponding ``walk operator" would then read in the state space of
(qumode)$\otimes$(coin)
\begin{eqnarray}
\hat{W} = \hat{\cal T}(\xi_d) \left(\hat{I} \otimes \hat{\cal C}\right) \, .
\label{W_op}
\end{eqnarray}
Here $\hat{I}$ is the qumode identity operator and $\hat{\cal T}$ is
a translation operator whose action is conditioned on the coin
configuration
\begin{eqnarray}
\hat{\cal T}(\xi_d) = \hat{D}(+\xi_d) \otimes |R\rangle\langle R|
+ \hat{D}(-\xi_d) \otimes |L\rangle\langle L| \, .
\label{T_op}
\end{eqnarray}
In order to prepare GKP qubits for the qumode, we shall consider 1D
QW of a position-squeezed vacuum state along with a coin qubit in an
arbitrary configuration
\begin{eqnarray}
|\psi_{\rm in}\rangle = |0\rangle_r \left(\, \alpha|R\rangle+\beta|L\rangle\, \right) \, ,
\label{psi_0}
\end{eqnarray}
where $|\alpha|^2+|\beta|^2=1$. As we will demonstrate below, by
means of 1D QW in phase space, one can transcribe the ``logical
state" imprinted in the coin configuration in \eref{psi_0} onto the
qumode ``coordinate" degrees of freedom $\{|q\rangle_r\}$. Since
different choices for the coin-toss transformation $\hat{\cal C}$
can lead to rather distinct walk patterns in the QW
\cite{Ke03,Ma14}, it can be anticipated that different encodings can
be achieved through different coin-toss transformations. In the
following, we will consider first the case of a ``Hadamard
coin-toss", which induces unitary evolution of the input state
\eref{psi_0}. As we shall find out, the consequent codewords will be
quite different from the approximate GKP codewords in
\eref{GKP_wfs}. We will then turn to another coin-toss
transformation, which will engender codeword states that are similar
to the ``standard" ones in \eref{GKP_wfs}.

\subsection{\label{sec:unit}Generic (unitary) quantum-walk encoding}
\begin{figure*}
\begin{center}
\includegraphics*[width=120mm]{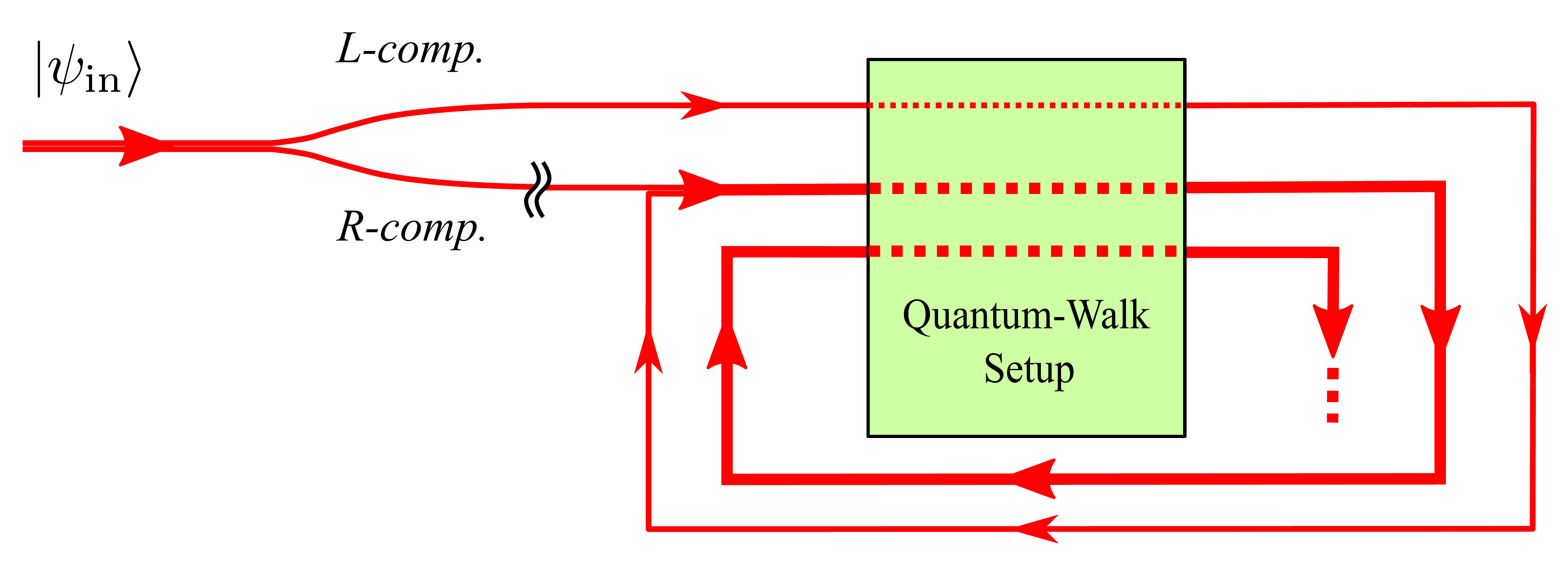}
\end{center}
\caption{Conceptual layout for our QW-encoding scheme, where each
thin line represents the respective component of the input state
$|\psi_{\rm in}\rangle$ [see Eq.~\eref{psi_0}] as indicated, while
heavy lines stand for the full state. Each passing (dashed line) of
the state through the QW-setup implements for the qumode one single
step of QW in phase space. Before entering the QW-setup, the input
state $|\psi_{\rm in}\rangle$ is split into its $R$ and
$L$-components, with the $R$-component subsequently delayed behind
the $L$-component. After the $L$-component has completed one single
step of QW while none for the $R$-component, the two are combined
for the ensuing QWs. The encoded state is generated after the
combined state has completed the designated steps of QW.
\label{fig:qwGKP}}
\end{figure*}

For a walker that localizes initially at the origin, after even
(odd) steps of QW the wavefunction of the walker would become
coherent superposition of localized states over even (odd) multiples
of the step length. In view of the structure of the ideal GKP
codewords \eref{ideal_GKP}, it is clear that one can exploit this
feature of QWs to prepare GKP qubits. The conceptual plan for our
encoding scheme is outlined in Fig.~\ref{fig:qwGKP}. In order to
simplify our theoretical formulation, let us suppose for now that
when implementing the QW, we are able to separate and combine the
components, e.g. $|0\rangle_r|R\rangle$ and $|0\rangle_r|L\rangle$
(which, for brevity, will be referred to as the $R$ and
$L$-components, respectively) of the input state \eref{psi_0} in the
way shown in Fig.~\ref{fig:qwGKP}. As we shall find out, this serves
only as a convenience to help present our theoretical ideas in a
clear way, but won't be a necessity in physical implementations for
the scheme, which we shall discuss in Sec.~\ref{sec:implm}.

In our encoding scheme, as shown in Fig.~\ref{fig:qwGKP}, prior to
the QW we separate the $R$ and the $L$-components of the input state
\eref{psi_0} to produce a time gap between them. In the scenario of
Fig.~\ref{fig:qwGKP}, we delay the $R$-component so that when it
begins its QW, the $L$-component would have completed exactly one
step of QW. The two components are then combined for all subsequent
QWs. Since the $R$-component always lags behind the $L$-component by
one single step, in the course of the QW coherent superposition of
localized spikes at sites of opposite parities are generated
progressively for the two components of the input state. Therefore,
a GKP-type encoding can be furnished after the state has completed
the desired number of steps of QW.

As an illustration, let us consider an encoding with the following
coin-toss transformation in the coin basis $\{|R\rangle,|L\rangle\}$
\begin{eqnarray}
\hat{\cal C}_H = \frac{1}{\sqrt{2}}
          \left(
            \begin{array}{cc}
              1 &  1 \\
              1 & -1 \\
            \end{array}
          \right) \, ,
\label{H-coin}
\end{eqnarray}
which corresponds to a Hadamard coin-toss for the QW \cite{Ma14}.
Suppose the initial state \eref{psi_0} undergoes the encoding
process of Fig.~\ref{fig:qwGKP}, so that its $R$ and $L$-components
would complete, respectively, $N$ and $(N+1)$ steps of QW upon
output. It then leads to the state
\begin{eqnarray}
|\psi_{\rm out}\rangle = \alpha |\psi_N^{(R)}\rangle + \beta |\psi_{N+1}^{(L)}\rangle \, ,
\label{psi_out}
\end{eqnarray}
where $|\psi_N^{(\epsilon)}\rangle$ with $\epsilon=\{R,L\}$ are the
resulting states for $|0\rangle_r|\epsilon\rangle$ after $N$ steps
of QW. One can find analytically that \cite{Am01}
\begin{eqnarray}
|\psi_N^{(\epsilon)}\rangle
= \sideset{}{'}\sum_{n=-N}^N \,\,|n\rangle_r \left(u_N^{(\epsilon)}(n) |R\rangle
+ v_N^{(\epsilon)}(n) |L\rangle \right) \, ,
\label{psi_N_form}
\end{eqnarray}
where $\sum_n'$ indicates summation over every other integers, that
is, $n=\{-N$, $-(N-2)$, $\dots$, $(N-2)$, $N\}$. To avoid
distractions, we supply explicit expressions for the amplitudes
$u_N^{(\epsilon)}(n)$ and $v_N^{(\epsilon)}(n)$ in Appendix
\ref{sec:uv1}. From \eref{psi_N_form}, it is then clear that when
$N$ is, for instance, even $|\psi_N^{(R)}\rangle$ in \eref{psi_out}
would cover only even sites, while $|\psi_{N+1}^{(L)}\rangle$ only
odd sites. By defining the encoded logical basis states here
\begin{eqnarray}
|0\rangle_{\rm QW}\equiv|\psi_N^{(R)}\rangle
\quad {\rm and}\quad
|1\rangle_{\rm QW}\equiv|\psi_{N+1}^{(L)}\rangle \, ,
\label{QW_code}
\end{eqnarray}
we then have from \eref{psi_out} the encoded state for the input
state \eref{psi_0}
\begin{eqnarray}
|\psi_{\rm encd}\rangle = \alpha |0\rangle_{\rm QW} + \beta |1\rangle_{\rm QW} \, .
\label{psi_N_unit}
\end{eqnarray}
Our scheme thus furnishes a GKP-type encoding for an arbitrary input
state.

\begin{figure}
\begin{center}
\includegraphics*[width=100mm]{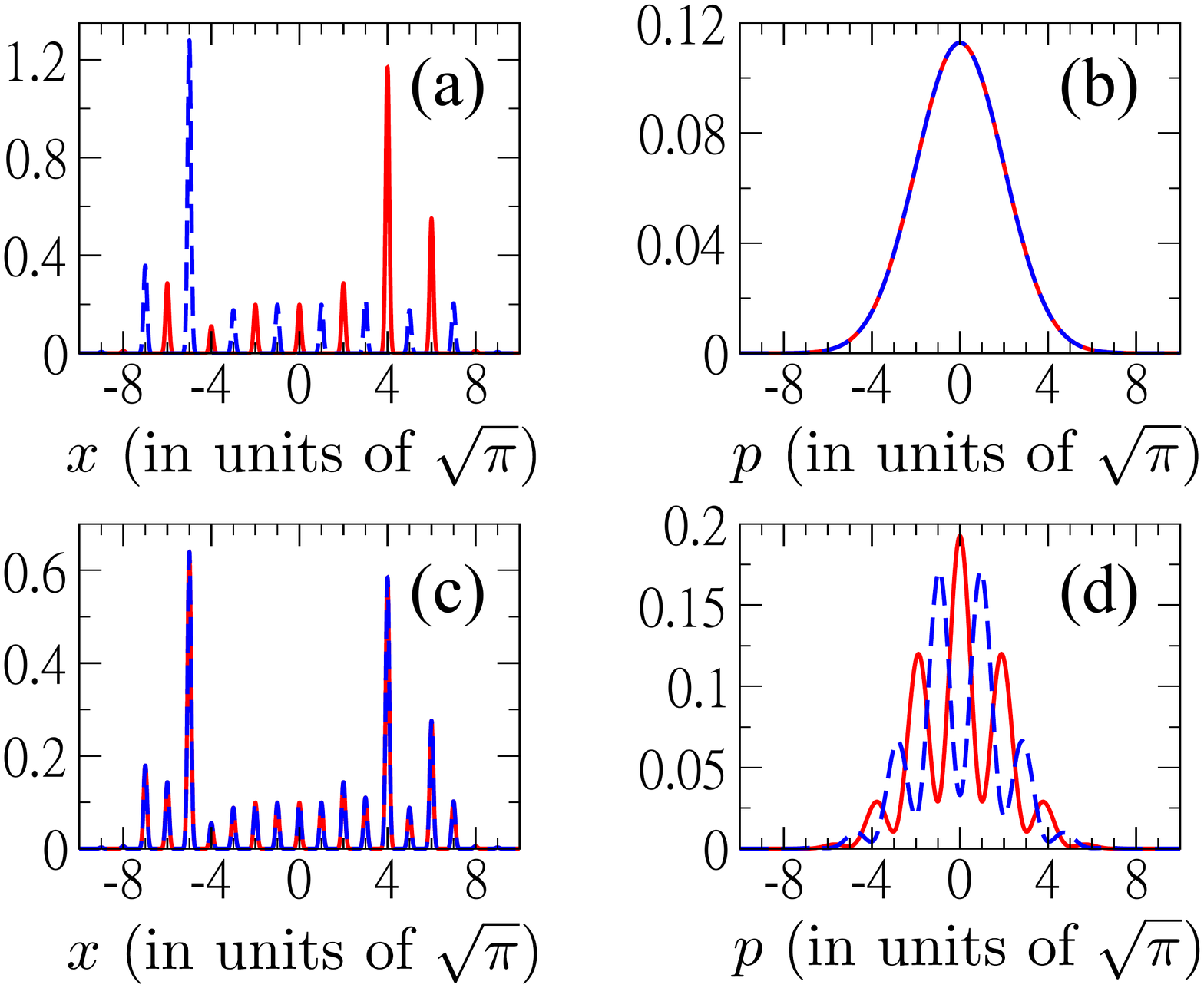}
\caption{Probability densities in position space and momentum space
for generic QW-codewords with $N=8$ at a squeezing with
$e^{-r}=0.2$ ($\sim$13.98 dB). Panels (a) and (b) illustrate
the results for $|0\rangle_{\rm QW}$ (solid curves) and $|1\rangle_{\rm QW}$
(dashed curves), while panels (c) and (d) show those
for $|+\rangle_{\rm QW}\equiv(|0\rangle_{\rm QW}+|1\rangle_{\rm QW})/\sqrt{2}$ (solid curves)
and $|-\rangle_{\rm QW}\equiv(|0\rangle_{\rm QW}-|1\rangle_{\rm QW})/\sqrt{2}$ (dashed curves).
\label{fig:generic}}
\end{center}
\end{figure}

To find the position-space and the momentum-space wavefunctions for
the codewords $|0\rangle_{\rm QW}$ and $|1\rangle_{\rm QW}$, we note
that from \eref{qx} one can obtain the wavefunctions for the
squeezed coherent states $|n\rangle_r$
\begin{eqnarray}
{}_x\langle x|n\rangle_r &=& \frac{e^{+\frac{r}{2}}}{\pi^{\frac{1}{4}}}
\exp\!\left[\frac{-(x-n\,x_d)^2}{2\,e^{-2r}}\right] \, ,
\nonumber \\
{}_p\langle p|n\rangle_r &=& \frac{e^{-\frac{r}{2}}}{\pi^{\frac{1}{4}}}
\exp\!\left[+i\,n\,x_d\,p-\frac{p^2}{2\,e^{+2r}}\right] \, ,
\label{n_wfs}
\end{eqnarray}
where $x_d\equiv\sqrt{2}\xi_d$, as before (see footnote
\ref{fn:sqrt2}). Making use of \eref{n_wfs}, one can obtain the
wavefunctions for the codeword states $|0\rangle_{\rm
QW}=|\psi_N^{(R)}\rangle$ and $|1\rangle_{\rm
QW}=|\psi_{N+1}^{(L)}\rangle$ using \eref{psi_N_form} and the
expressions for the amplitudes $u_N^{(\epsilon)}$,
$v_N^{(\epsilon)}$, and etc.~in Appendix \ref{sec:uv1}. Setting
$x_d=\sqrt{\pi}$ for the QW, we show in Fig.~\ref{fig:generic} the
probability densities in position and momentum spaces for the
QW-codewords $|0\rangle_{\rm QW}$ and $|1\rangle_{\rm QW}$, together
with the conjugate states $|+\rangle_{\rm QW}\equiv(|0\rangle_{\rm
QW}+|1\rangle_{\rm QW})/\sqrt{2}$ and $|-\rangle_{\rm
QW}\equiv(|0\rangle_{\rm QW}-|1\rangle_{\rm QW})/\sqrt{2}$ for the
case of $N=8$ at a squeezing with $e^{-r}=0.2$ (corresponding to
$\sim$13.98 dB squeezing).\footnote{Although probability densities
must be normalized to (i.e.,~integrated to yield) the value 1.0,
there can still be sharp peaks greater than 1.0 provided they are
sufficiently narrow, such as those in Fig.~\ref{fig:generic}(a).}

From Fig.~\ref{fig:generic}, we see that, despite the unusual
envelopes, the probability densities of the QW-codewords exhibit
features characteristic of approximate GKP codewords that are
essential for correcting shift errors \cite{Pi04}. In particular,
the position distributions of the codewords $|0\rangle_{\rm QW}$ and
$|1\rangle_{\rm QW}$ consist of Gaussian spikes at, respectively,
even and odd multiples of $\sqrt{\pi}$, while the momentum
distributions of the codewords $|+\rangle_{\rm QW}$ and
$|-\rangle_{\rm QW}$ manifest peaks at, respectively, even and odd
multiples of $\sqrt{\pi}$. Therefore, in principle, the QW-codewords
here can be adopted to correct shift errors in accordance with the
GKP scheme, despite the issues with their performance and probably
also efficiencies \cite{Me14}. The key drawbacks of the QW-codewords
\eref{QW_code} reside in their momentum distributions, such as the
featureless Gaussians in Fig.~\ref{fig:generic}(b) for
$|0\rangle_{\rm QW}$ and $|1\rangle_{\rm QW}$, and the broad peaks
in Fig.~\ref{fig:generic}(d) for $|+\rangle_{\rm QW}$ and
$|-\rangle_{\rm QW}$, which would render the correction for
$p$-errors ineffective. We find numerically that these features are
independent of the QW steps $N$, likely due to unitarity of the QW
here. Since different coin-toss transformations for QW can lead to
very different walk patterns \cite{Ke03,Ma14}, one possible remedy
for the present dilemma is to replace the coin-toss operation
\eref{H-coin} with a new one. Alternatively, since the difficulty
with the codewords $|0\rangle_{\rm QW}$ and $|1\rangle_{\rm QW}$
lies in the lack of (coin-state independent) structures in their
momentum distributions, one can attempt to implement periodic
structures in both position and momentum directions through
two-dimensional QWs \cite{Ma14}. As our point here is to demonstrate
the feasibility of the proposed QW-scheme for generating GKP-type
encodings in a generic setting, we shall not pursue these issues
further.

In addition to the ``unconventional" profiles in the probability
densities of the QW-codewords in Fig.~\ref{fig:generic}, it should
be noted that here the codewords $|0\rangle_{\rm QW}$ and
$|1\rangle_{\rm QW}$ are entangled states between the qumode and the
ancillary coin-qubit [see \eref{psi_N_form}]. This is in stark
contrast to codewords from the original GKP-scheme, where the
encoding resides entirely in the qumode. To disentangle the qumode
from the coin qubit in the QW-codewords, one can choose to project
them onto any given coin state, such as the symmetric ``diagonal"
coin-state $|D\rangle\equiv (|R\rangle+|L\rangle)/\sqrt{2}$.
However, we find this would lead to states that are plagued by fast
oscillations of intervals below $\sqrt{\pi}$ in their momentum-space
distributions, which are unfavorable for correcting $p$-errors.
Therefore, here we choose to retain the form \eref{psi_N_form} for
the QW-codewords with its full generality. At this point, it is then
natural to ask whether our scheme is capable of producing
QW-codewords akin to the ``standard" GKP ones as in \eref{GKP_wfs}
or not. As we will now show, this is indeed possible if appropriate
coin-toss transformation is used.

\subsection{\label{sec:dissp}Dissipative (non-unitary) quantum-walk encoding}
In order to generate codewords similar to \eref{GKP_wfs}, it
necessitates implementing QWs with Gaussian probability
distributions in our scheme. Intuitively, one might expect
decoherence of the state must be incorporated, so that the QW would
become classical and yield the desired probability distributions
\cite{Ke07}. However, this would inevitably lead to mixed states,
which are unfavorable for our purposes here, as the codeword states
must be pure states. To find the way out, we note that the
nonclassical nature of the QW arises from the interference between
the coin-toss outcomes for the $R$ and the $L$-components.
Therefore, if we project the state vector at each step of the QW, so
that the two coin-state components won't interfere, it would then be
possible to generate coherent superposition of ``classically"
distributed Gaussian spikes. In other words, by ``resetting" the
coin state of the walker to a symmetric combination of the
$|R\rangle$ and the $|L\rangle$ states in each step of the QW, one
can then generate the desired walk pattern here. It thus follows
that one should replace the coin-toss transformation \eref{H-coin}
with the projection operator for the diagonal coin-state
$|D\rangle=(|R\rangle+|L\rangle)/\sqrt{2}$, i.e.,
\begin{eqnarray}
\hat{\cal C}_D = |D\rangle\langle D| =
\frac{1}{2}
          \left(
            \begin{array}{cc}
              1 & 1 \\
              1 & 1 \\
            \end{array}
          \right) \, .
\label{HD-coin}
\end{eqnarray}
With this change, as we shall show below, we are then able to
achieve the targeted codeword states. Since the projection operator
can reduce the total probability of the state it acts on, the QW
here becomes nonunitary, and we shall refer to it as the
``dissipative" QW.

For the QW-encoding modified with the new coin-toss operation
\eref{HD-coin}, the calculation for the corresponding encoded states
proceeds in exactly the same manner as before. After the
$R$-component and the $L$-component of the input state \eref{psi_0}
have completed, respectively, $N$ and $(N+1)$ steps of QW, one can
find an output state with the same structure as Eqs.~\eref{psi_out}
and \eref{psi_N_form}, but now with different explicit forms for the
amplitudes $u_N^{(\epsilon)}(n)$ and $v_N^{(\epsilon)}(n)$ (see
Appendix \ref{sec:uv2}). At this point, it is tempting to conclude
immediately that the encoding can then be done in exactly the same
way as in \eref{QW_code}. This is, however, incorrect because we now
have nonunitary, dissipative QWs and thus must take extra care for
the normalization of the state vectors. Moreover, in order that the
codeword states would resemble better the approximate GKP codewords
\eref{GKP_wfs}, we find it advantageous to project the final state
of the QW onto the diagonal coin-state $|D\rangle$, which also
serves to disentangle the qumode from the coin qubit here. For the
initial state $|0\rangle_r|\epsilon\rangle$ the resulting {\em
unnormalized} state vector after $N$ steps of dissipative QW
(including the action of the $|D\rangle$-projector upon output)
takes the form
\begin{eqnarray}
|\psi_N\rangle =
\sideset{}{'}\sum_{n=-N}^N\!w_N(n)\,\,|n\rangle_r |D\rangle \, ,
\label{psi_N_dQW1}
\end{eqnarray}
where
\begin{eqnarray}
w_N(n) = \frac{1}{2^{N+\frac{1}{2}}}
\left(
  \begin{array}{c}
     N \\
     \frac{N+n}{2}\\
  \end{array}
\right) \, .
\label{wN1}
\end{eqnarray}
Notice that here the qumode and the coin qubit are fully
disentangled. Also, since the $R$ and the $L$-components are now
symmetrical, we have dropped the superscript $\epsilon$ for the
initial coin configurations in \eref{psi_N_dQW1} and \eref{wN1}
[cf.~Eq.~\eref{psi_N_form} for the unitary case]. In terms of the
normalized state vectors for \eref{psi_N_dQW1}
\begin{eqnarray}
|\phi_N\rangle \equiv Z_N^{-1/2}|\psi_N\rangle
\label{psi_N_dQW}
\end{eqnarray}
with $Z_N\equiv\langle\psi_N|\psi_N\rangle$, we find the output
state for the dissipative QW
\begin{eqnarray}
|\psi_{\rm out}\rangle &=& \alpha\,\sqrt{Z_N}\,\,|\phi_N\rangle
+ \beta\,\sqrt{Z_{N+1}}\,\,|\phi_{N+1}\rangle
\nonumber \\
&\propto& \alpha' |\phi_N\rangle + \beta'|\phi_{N+1}\rangle \, ,
\label{psi_out_dQW}
\end{eqnarray}
where we have denoted in the second line
$\alpha'\equiv\alpha/\sqrt{|\alpha|^2 + \gamma^2 |\beta|^2}$ and
$\beta'\equiv\gamma\beta/\sqrt{|\alpha|^2 + \gamma^2 |\beta|^2}$
with $\gamma\equiv \sqrt{Z_{N+1}/Z_N}$. Therefore, identifying
\begin{eqnarray}
|0\rangle_{\rm dQW}\equiv|\phi_N\rangle
\quad {\rm and}\quad
|1\rangle_{\rm dQW}\equiv|\phi_{N+1}\rangle \, ,
\label{dQW_code}
\end{eqnarray}
we arrive at the following encoding for the input state \eref{psi_0}
\begin{eqnarray}
|\psi_{\rm encd}\rangle = {\cal N}\,(\,\alpha' |0\rangle_{\rm dQW} + \beta' |1\rangle_{\rm dQW}\,)
\label{psi_encd_dQW}
\end{eqnarray}
with ${\cal N}\equiv\sqrt{|\alpha|^2 Z_N + |\beta|^2 Z_{N+1}}$.
Notice that the coefficients $\alpha$, $\beta$ in the original state
\eref{psi_0} have been modified in the final encoding
\eref{psi_encd_dQW} due to the dissipative nature of the QW.
Therefore, when applying this encoding scheme, one must prepare the
input state properly, so that the desired encoded states can be
obtained at the output.

\begin{figure}
\begin{center}
\vspace*{-40mm}
\includegraphics*[width=150mm]{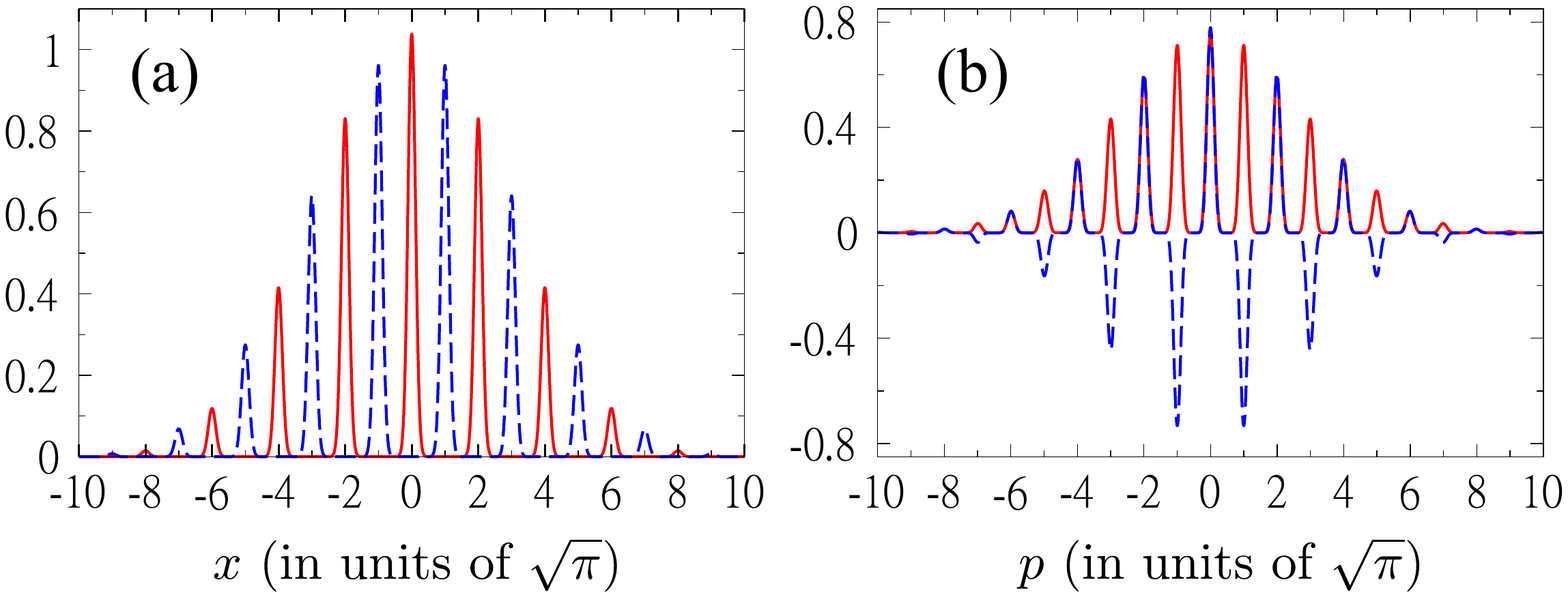}
\caption{(a) Position and (b) momentum wavefunctions for the
dissipative QW-codewords $|0\rangle_{\rm dQW}$ (solid curves)
and $|1\rangle_{\rm dQW}$ (dashed curves) with $N=8$ at a squeezing with
$e^{-r}=1/\sqrt{8\pi}\sim 0.199$ ($\sim 14.00$ dB squeezing).
\label{fig:dissp}}
\end{center}
\end{figure}

To find the wavefunctions of the codewords here, one can again use
\eref{n_wfs} in \eref{dQW_code} [together with
Eqs.~\eref{psi_N_dQW1}--\eref{psi_N_dQW}]. For the momentum-space
wavefunction, the summation over the site index $n$ can be done
analytically. We find
\begin{eqnarray}
\hspace*{-5mm}
{}_p\langle p|\phi_N\rangle = \left(\frac{e^{-r}}{2\sqrt{\pi}Z_N}\right)^{\!\!\frac{1}{2}}
\!\!\exp\!\!\left[-\frac{p^2}{2\,e^{+2r}}\right]\!\cos^N\!(x_d\,p) |D\rangle\!.
\label{p_wf_dQW}
\end{eqnarray}
For the position-space wavefunction, it is of particular interest to
examine its large $N$ limit, for which the binomial distribution
would tend to a Gaussian. Applying Stirling's formula, we find for
large $N$
\begin{eqnarray}
{}_x\langle x|\phi_N\rangle \approx
\left(\frac{2e^{+r}}{\pi\sqrt{N}}\right)^{\!\!\frac{1}{2}}
\!\!\!
\sideset{}{'}\sum_{n=-N}^N\!e^{-\frac{n^2}{2N}} \,
\exp\!\!\left[-\frac{(x-n\,x_d)^2}{2\,e^{-2r}}\right] |D\rangle \,.
\nonumber \\
\label{q_wf_dQW}
\end{eqnarray}
For the normalization in \eref{q_wf_dQW}, we have taken the squeezed
coherent states $\{|n\rangle_r\}$ here to be approximately
orthogonal, so that $Z_N$ in \eref{psi_N_dQW} can be approximated as
\begin{eqnarray}
Z_N \approx \frac{1}{2^{2N+1}}
\left(
  \begin{array}{c}
     2N \\
     N  \\
  \end{array}
\right)
\approx \frac{1}{2\sqrt{\pi N}} \, .
\label{ZN_approx}
\end{eqnarray}
Taking $x_d=\sqrt{\pi}$ in \eref{q_wf_dQW} and comparing the
expression with the approximate GKP codewords \eref{GKP_wfs}, we see
that in the large $N$ limit the dissipative QW (or dQW, for short)
codewords \eref{dQW_code} correspond to approximate GKP codewords
with width $\Delta_x \approx e^{-r}$ and $\Delta_p \approx
1/\sqrt{\pi N}$. Therefore, for shift errors symmetric in the
position and the momentum quadratures, following GKP \cite{GKP01},
the choice for encodings with $\Delta_x=\Delta_p$ becomes here
$e^{+r}=\sqrt{N\pi}$. As an illustration, we plot in
Fig.~\ref{fig:dissp} the wavefunctions for the dQW-codewords for the
case with $N=8$ and $e^{-r}=1/\sqrt{8\pi}\sim 0.199$ (corresponding
to $\sim 14.00$ dB squeezing), which carry the hallmarks of
approximate GKP codewords \eref{GKP_wfs}.

\begin{figure}
\begin{center}
\vspace*{-15mm}
\includegraphics*[width=90mm]{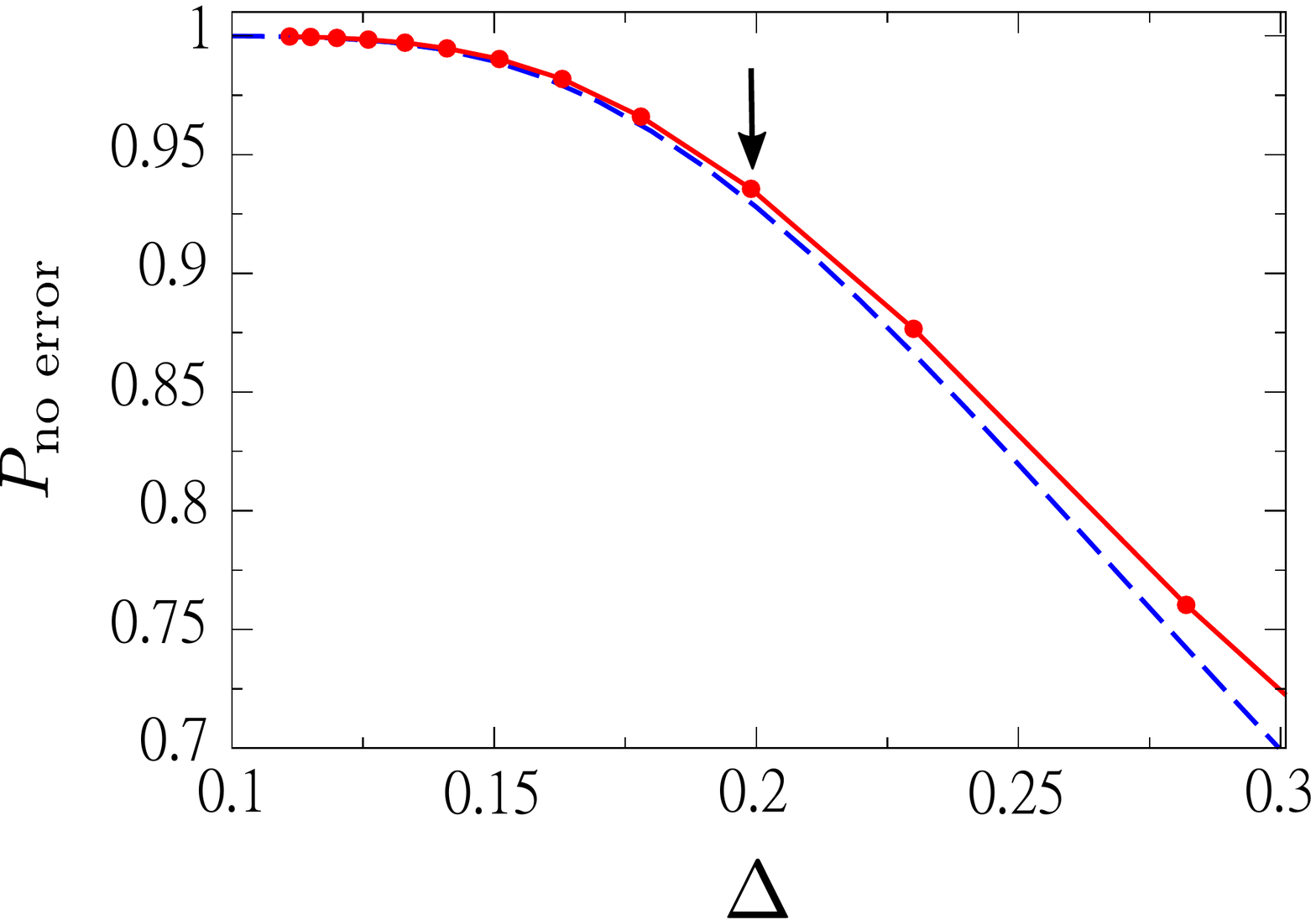}
\caption{The probability for repeated error corrections without Pauli
errors using dissipative QW-codeword $|0\rangle_{\rm dQW}$ (dots) with
$\Delta=e^{-r}=1/\sqrt{N\pi}$,
where the solid line serves as guide to the eyes. For comparison, we also plot the
results for the approximate GKP codeword $|\tilde{0}\rangle_L$ (dashed line) with
$\Delta_x=\Delta_p=\Delta$ in \eref{GKP_wfs}. The
arrow indicates the point for $N=8$ and $\Delta=1/\sqrt{8\pi}\sim 0.199$
(corresponding to $\sim 14.0$ dB squeezing) in the dissipative QW-codeword.
\label{fig:Pnerr}}
\end{center}
\end{figure}

In order to evaluate the performance of the dQW-codewords
\eref{dQW_code}, we consider the error-correcting scheme of
Ref.~\cite{Gl06} and find the probability $P_{\mbox{\scriptsize no
error}}$ for repeated error corrections using the dQW-codewords
without incurring Pauli errors (see Appendix \ref{sec:Pnerr} for
details).\footnote{ Of course, the figure of merit
$P_{\mbox{\scriptsize no error}}$ here is by no means the unique
measure for the performance of our codewords. For instance, if one
is concerned with the codewords' resilience to photon loss, one then
has to resort to other measures; see, e.g., Ref.~\cite{Al18}. Here
we are checking the performance of our codewords for correcting
shift errors, which is what the GKP encoding was proposed to tackle
originally, the Glancy-Knill scheme \cite{Gl06} thus provides an
appropriate means for our purposes. } For this calculation, we
consider $N$-step dissipative QW-encoding with width
$e^{-r}=1/\sqrt{N\pi}\equiv\Delta$. The results are shown in
Fig.~\ref{fig:Pnerr}, where we also plot the results for the
approximate GKP codewords \eref{GKP_wfs} with
$\Delta_x=\Delta_p=\Delta$ for comparison. It is encouraging to find
that the dQW-codewords in fact outperform their GKP counterparts for
all $\Delta$ by a small margin. In particular, for the $N=8$
dQW-codewords we get $P_{\mbox{\scriptsize no error}}\simeq 0.936$
[vs. $\simeq 0.929$ for the GKP case] at the squeezing $\sim 14.00$
dB, which is within current experimental capabilities \cite{Va16}.
Although for dQW-codewords with $N=10$ one can even attain
$P_{\mbox{\scriptsize no error}}\simeq 0.966$ at a squeezing $\sim
14.97$ dB, which lies barely below the 15 dB squeezing achieved in
Ref.~\cite{Va16}, implementing 10 and 11 steps of QW are even more
challenging experimentally.

\section{\label{sec:implm}Implementations}
\begin{figure*}
\begin{center}
\includegraphics*[width=100mm]{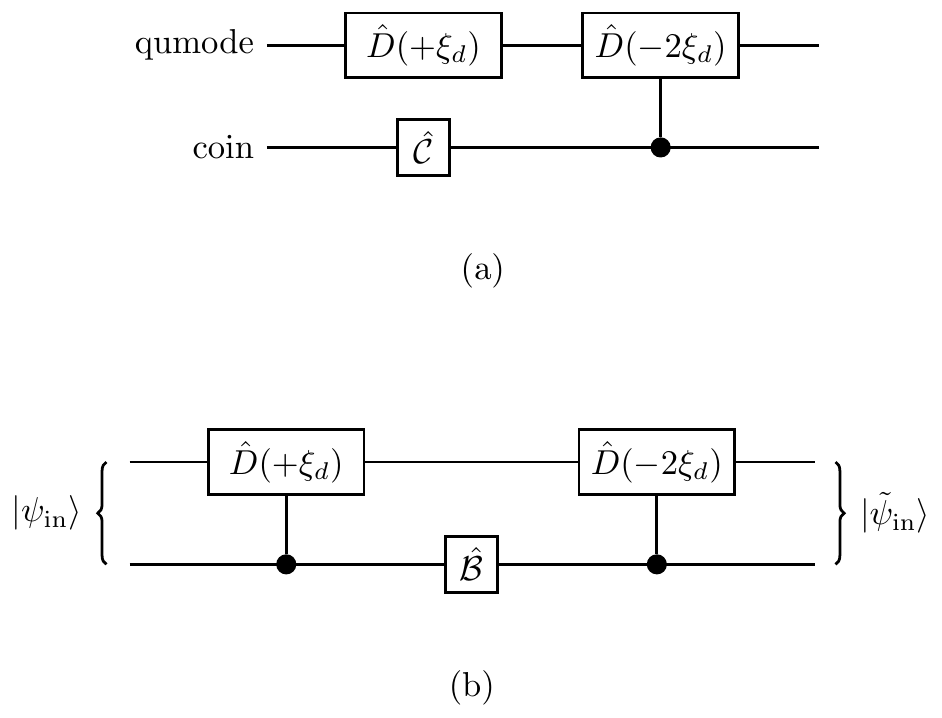}
\caption{Quantum circuits for (a) implementing one single step of QW
and (b) preparing the properly delayed state $|\tilde{\psi}_{\rm
in}\rangle$ from the input state $|\psi_{\rm in}\rangle$. Here
$\hat{D}$ is the phase-space displacement operator, $\hat{\cal C}$
the coin-toss operator for the QW, and $\hat{\cal B}$ the ``biased"
coin-toss operator for state preparation (see text).
\label{fig:QWcir}}
\end{center}
\end{figure*}

We shall now look into the physical implementations for our
QW-encoding scheme. To begin with, let us take the coin states
$|R\rangle$ and $|L\rangle$, respectively, the logical zero and the
logical one states of a control qubit for a controlled-displacement
gate, and write the walk operator $\hat{W}$ of \eref{W_op} as
\begin{eqnarray}
\hat{W} = \left[ \hat{I} \otimes |R\rangle\langle R|
+ \hat{D}(-2\xi_d) \otimes |L\rangle\langle L| \right]\,
\left(\hat{D}(+\xi_d) \otimes \hat{\cal C}\right) \, .
\label{W_op_impl}
\end{eqnarray}
It then follows that the walk operator $\hat{W}$ can be implemented
through the quantum circuit depicted in Fig.~\ref{fig:QWcir}(a). To
prepare logical basis states in the QW-scheme [i.e., \eref{QW_code}
or \eref{dQW_code}], one can thus supply the states
$|0\rangle_r|R\rangle$ and $|0\rangle_r|L\rangle$ to the circuit
separately and cycle for the corresponding numbers of rounds for the
QW. It is to be noted that for the dissipative QW-encoding of
Sec.~\ref{sec:dissp}, an additional projection operator
$|D\rangle\langle D|$ over the coin degree of freedom has to be
applied at the output of the final round of QW [see above
Eq.~\eref{psi_N_dQW1}], which is not included in the circuit of
Fig.~\ref{fig:QWcir}(a). To encode qubits with arbitrary logic,
however, it requires additional efforts, as one has to delay either
the $R$ or the $L$-component of the general input state \eref{psi_0}
for the encoding.

As previously, let us delay the $R$-component of the input state one
step behind its $L$-component for the QW. It then suffices if we are
able to prepare from the input state $|\psi_{\rm in}\rangle$ the
corresponding ``delayed state" $|\tilde{\psi}_{\rm in}\rangle$,
which has the $L$-component already taken one step of QW, while none
for the $R$-component, that is,
\begin{eqnarray}
|\tilde{\psi}_{\rm in}\rangle &=& \alpha |0\rangle_r|R\rangle
+ \beta\,\hat{W}\,|0\rangle_r|L\rangle \, .
\label{psi_dely}
\end{eqnarray}
For instance, in the case of the unitary QW-encoding with a Hadamard
coin \eref{H-coin} discussed in Sec.~\ref{sec:unit}, the desired
delayed state would then be
\begin{eqnarray}
|\tilde{\psi}_{\rm in}\rangle_{\rm QW}
= \alpha |0\rangle_r|R\rangle + \frac{\beta}{\sqrt{2}}
\left( |+1\rangle_r|R\rangle - |-1\rangle_r|L\rangle\right) \, .
\label{psi_dely_H}
\end{eqnarray}
As shown in Appendix \ref{sec:statprep}, this state can be produced
from the input state $|\psi_{\rm in}\rangle$ making use of the
circuit illustrated in Fig.~\ref{fig:QWcir}(b), which has two
controlled-displacement operators separated by a ``biased" coin-toss
operator $\hat{\cal B}=\hat{\cal B}_H$ given by
\begin{eqnarray}
\hat{\cal B}_H =
          \left(
            \begin{array}{cc}
              1 & 0 \\
              \frac{1}{\sqrt{2}} & \frac{-1}{\sqrt{2}} \\
            \end{array}
          \right) \, .
\label{B_H}
\end{eqnarray}
Comparing \eref{B_H} with \eref{H-coin}, we see that $\hat{\cal
B}_H$ does not flip the $R$-component of the quantum coin, while
tosses its $L$-component in the way of the Hadamard coin $\hat{\cal
C}_H$ (thus, a ``biased" coin-toss). It is therefore not surprising
that the delayed state can be duly prepared this way.

Similarly, for the dissipative QW-encoding of Sec.~\ref{sec:dissp}
the necessary delayed state can be obtained using \eref{psi_dely}
and the corresponding walk operator. The result reads
\begin{eqnarray}
|\tilde{\psi}_{\rm in}\rangle_{\rm dQW} = \alpha |0\rangle_r|R\rangle + \frac{\beta}{2}
\left( |+1\rangle_r|R\rangle + |-1\rangle_r|L\rangle\right) \, .
\label{psi_dely_dissp}
\end{eqnarray}
Again, this state can be prepared through the circuit of
Fig.~\ref{fig:QWcir}(b) with the following biased coin-toss
\begin{eqnarray}
\hat{\cal B}_D =
          \left(
            \begin{array}{cc}
              1 & 0 \\
              \frac{1}{2} & \frac{1}{2} \\
            \end{array}
          \right) \, .
\label{B_D}
\end{eqnarray}
Once the delayed state is available, for both encoded states
\eref{psi_N_unit} and \eref{psi_encd_dQW} discussed in
Sec.~\ref{sec:form}, the remaining $N$ steps of QW can then be
implemented by sending the respective $|\tilde{\psi}_{\rm
in}\rangle$ into the circuit of Fig.~\ref{fig:QWcir}(a) and cycling
for $N$ rounds. As before, in the dissipative case an additional
projector $|D\rangle\langle D|$ over the coin qubit must be
incorporated into the circuit of Fig.~\ref{fig:QWcir}(a) at the
output of the final round.

With the architecture for implementing QW-encodings in place, it is
of interest to examine how it can be realized in practice. One
possible route that may be accessible to current technologies is
offered by extending the circuit quantum-electrodynamics (cQED)
setup for implementing the cat code \cite{Co99,Le13,Vl13,Of16}. To
realize our encoding scheme, we see from Fig.~\ref{fig:QWcir} that
it calls for implementing conditional and unconditional qumode
phase-space displacements, along with coin operations with biased
($\hat{\cal B}$) and fair ($\hat{\cal C}$) coin-toss. As we shall
explain below, these gate operations can be implemented efficiently
using cQED systems, as demonstrated previously in the cat code and
other experiments.

\begin{figure*}
\begin{center}
\includegraphics*[width=90mm]{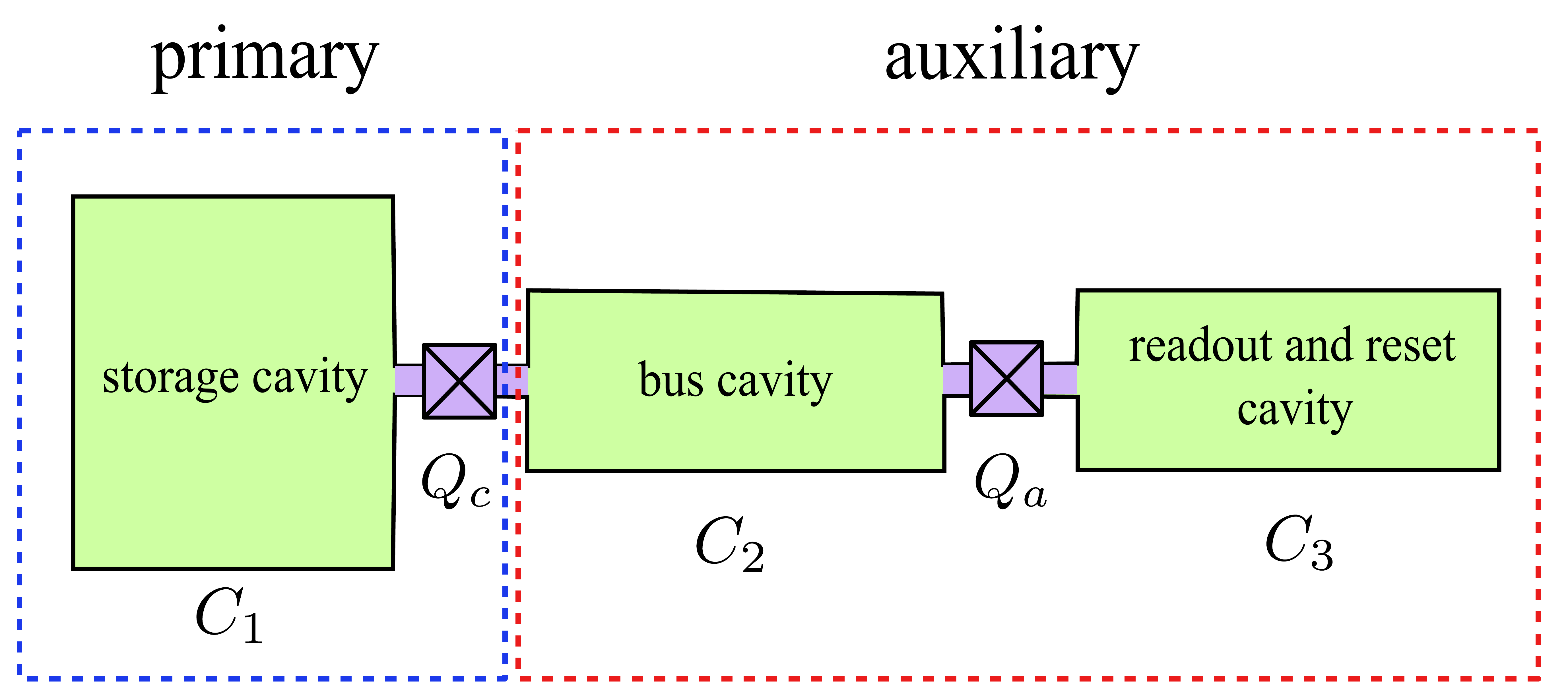}
\caption{Schematics of the proposed cQED setup for implementing the
QW-encoding, which is separated into the primary and the auxiliary
parts. The primary part consists of the main components for the
encoding, where a storage cavity ($C_1$) stores the codeword state
and a transmon qubit ($Q_c$) acts as the quantum coin for the QW. In
order to implement coin-toss operations, the auxiliary part contains
an ancilla qubit ($Q_a$) that is coupled to the coin qubit ($Q_c$)
through a bus cavity ($C_2$). Also in the auxiliary part is another
cavity ($C_3$) attached to the ancilla qubit ($Q_a$), which serves
to read out and reset the ancilla states at different stages of the
encoding. \label{fig:cqed}}
\end{center}
\end{figure*}

Let us consider the setup depicted schematically in
Fig.~\ref{fig:cqed}, which consists of three waveguide cavity
resonators $C_1\sim C_3$ intercepted with two superconducting
transmon qubits $Q_c$ and $Q_a$. For clarity, we have divided the
setup into two parts: the primary and the auxiliary parts. Here the
primary part consists of cavity $C_1$ for storing the codeword state
during the encoding process, and qubit $Q_c$ playing the role of the
quantum coin for the QW. The auxiliary part of the setup here serves
to furnish the coin-toss transformations required for the encoding,
including the biased one ($\hat{\cal B}$) for state preparation and
the fair one ($\hat{\cal C}$) for each step of the QW. It is
comprised of an ancillary qubit $Q_a$ connected to a bus cavity
$C_2$ for mediating coupling with the coin qubit $Q_c$, together
with cavity $C_3$ for measuring and resetting the ancilla state
during the QW.

To understand how the setup works, let us recall that generically
the interaction between a transmon qubit and a cavity mode is well
approximated by the Jaynes-Cummings Hamiltonian
\cite{Bl04,Gi14,We17}
\begin{eqnarray}
\frac{\hat{H}_{\rm JC}}{\hbar} = -\frac{\omega_q}{2} \hat{Z} + \omega_r \hat{a}^\dagger\hat{a} +
g \left(\,\hat{a}^\dagger\otimes\,|0\rangle\langle 1| + \hat{a}\otimes|1\rangle\langle 0|\,\right) \, .
\label{H_JC}
\end{eqnarray}
Here $\hat{Z}\equiv \left(|0\rangle\langle 0|-|1\rangle\langle
1|\right)$ is the Pauli-$Z$ operator for the qubit with $|0\rangle$
and $|1\rangle$, respectively, the lower and upper qubit levels that
are separated by frequency $\omega_q$ apart. $\hat{a}$ is the mode
operator for the cavity mode with frequency $\omega_r$, and $g$
characterizes the qubit-cavity interaction strength. In the regime
of large detuning, $\delta\equiv(\omega_q-\omega_r)\gg g$ (the
strong dispersive regime), one can derive perturbatively an
effective Hamiltonian from \eref{H_JC} \cite{Gi14,Sc01}
\begin{eqnarray}
\frac{\hat{H}}{\hbar} = -\frac{(\omega_q+\chi)}{2} \hat{Z}
+ \omega_r \hat{a}^\dagger\hat{a} - \chi \,\hat{a}^\dagger\hat{a}\otimes\hat{Z} \, ,
\label{H_disp}
\end{eqnarray}
where $\chi\equiv g^2/\delta$ is the qubit frequency shift (the
``Lamb shift") due to the cavity coupling. It is the effective
qubit-cavity coupling $\hat{V}_I\equiv - \chi
\,\hat{a}^\dagger\hat{a}\otimes\hat{Z}$ in \eref{H_disp} that
enables efficient control and manipulations of qubit and cavity
states in this strong dispersive regime \cite{Bl04,Ha06}. In
particular, time evolution due to the coupling $\hat{V}_I$ induces a
qubit-state dependent phase-space rotation of the cavity mode, which
yields for a time duration $t$
\begin{eqnarray}
e^{-i\chi t\,\hat{a}^\dagger\hat{a}\otimes\hat{Z}} =
e^{-i\chi t\,\hat{a}^\dagger\hat{a}} \otimes |0\rangle\langle 0|
+e^{+i\chi t\,\hat{a}^\dagger\hat{a}} \otimes |1\rangle\langle 1| \, .
\label{cond_rot}
\end{eqnarray}
As detailed in Appendix \ref{sec:cond_displ}, the evolution operator
\eref{cond_rot} can help implement the conditional cavity
displacement operation that is key to the QW-encoding (see
Fig.~\ref{fig:QWcir}). At the same time, since $\hat{V}_I$ commutes
with both the qubit and the cavity Hamiltonians, it allows quantum
non-demolition measurement of the qubit state or the cavity state
\cite{Ha06}. In our case, this is utilized in the setup of
Fig.~\ref{fig:cqed} to read out the ancilla ($Q_a$) state through
its coupling with cavity $C_3$.

Extending the consideration above to the case of two transmon qubits
jointly coupled to a cavity bus resonator, the off-resonant coupling
can then mediate an indirect interaction between the pair of qubits
\cite{Bl04,Ma07}. If higher qubit excitations are taken into
account, they can be exploited to furnish two-qubit gates, such as
the conditional phase gate \cite{Di09}\footnote{See Eq.~\eref{U_phi}
in Appendix \ref{sec:biased_coin}.} that is helpful for implementing
the coin-toss transformations for the QW-encoding--the task of the
auxiliary components in Fig.~\ref{fig:cqed}. For example, to effect
the biased coin-toss operation $\hat{\cal B}$ on the coin qubit, it
can be done by realizing a suitable positive operator-valued measure
(POVM) for this operation. To this end, it is necessary to introduce
an ancilla qubit coupled to the coin qubit and enact two-qubit
controlled gates along with single-qubit gates. Subsequent ancilla
measurement would then signal success/failure of the gate operation.
In the auxiliary part of the setup in Fig.~\ref{fig:cqed}, the bus
cavity ($C_2$) serves to mediate the desired two-qubit coupling
between the coin qubit and the ancilla qubit, and the readout cavity
($C_3$) provides the means for detecting the ancilla state. It
should be noted that the ancilla qubit must be reset to its initial
configuration after each run of the coin-toss, which can be done
through feedback controlled pulses via cavity $C_3$ \cite{Of16}. In
Appendix \ref{sec:cqed_appdx}, we demonstrate the scheme for
realizing the coin-toss operations in the dissipative QW-encoding,
where both the biased and the fair coin-toss operations will be
considered.

It should be noted that our discussions above for the cQED
implementation have entirely left out nonlinearity effects inherent
in such systems. In a more quantitative analysis, these must be
taken into account properly. Also, we point out that our
implementations for the coin-toss operations for the dissipative
QW-encoding demonstrated in Appendix \ref{sec:cqed_appdx} are
non-deterministic. With increasing number of QW steps $N$, the
overall success rate for the QW-encoding drops exponentially $\sim
1/2^N$. This can also be anticipated from the form of the (fair)
coin-toss operator $\hat{\cal C}_D$ given by \eref{HD-coin}, which
projects the state progressively in the course of the QW, reflecting
the ``dissipative" nature of the QW. Thus, in practice, the
dissipative QW-encoding would work only for small numbers of $N$. In
spite of this, since the recently achieved first experimental
realization for the GKP codewords \cite{Fl19} corresponds to $N\sim
2$ in our case, we think the dissipative QW-encoding should be still
of experimental interest. Moreover, we use the dissipative
QW-encoding to demonstrate how the QW-encoding scheme can be adapted
to generate approximate GKP codewords akin to the ``standard" forms
\eref{GKP_wfs}. Ultimately, we wish to devise a robust unitary
QW-encoding (or a unitary encoding process for general QW-codewords)
which would alleviate the loss problem. Nevertheless, it should be
noted that the preparation for the ``delayed states" \eref{psi_dely}
in our QW-encoding is always nonunitary. Even for a unitary (fair)
coin, this state preparation has a weak nonunitarity due to the
finite overlap between the squeezed coherent states $|0\rangle_r$
and $|+1\rangle_r$. In practice, however, this nonunitarity would be
rather small and for QW-encoding with a unitary coin one can safely
take it to be unitary to good approximations.\footnote{For instance,
in the case of Fig.~\ref{fig:generic} we have $\xi_d=\sqrt{\pi/2}$
and $e^{-r}=0.2$. It then follows from \eref{qx} that ${}_r\langle
0|+1\rangle_r=\exp\{-\xi_d^2\, e^{+2r}/2\}\sim 10^{-9}$, which is
vanishingly small.}

Finally, we remark that in preparing GKP-type codeword states from
the squeezed vacuum the process must be non-Gaussian \cite{GKP01}.
From our discussions for the cQED implementation of the QW-encoding,
we see that by coupling the qumode to a quantum coin, one can enact
effective non-Gaussian evolution of the qumode, which underlies the
codewords' robustness against shift errors. In order to improve the
robustness of the codewords, as exemplified by the dissipative
QW-encoding, it seems nonunitary coin operation can be advantageous
for enhancing nonlinearity in the qumode dynamics. There thus seems
to be a trade-off between unitarity and codeword robustness in the
QW-encoding: Unitarity is favored for preserving codeword intensity,
but not for promoting its robustness. To achieve robust GKP-type
codeword states with optimum intensities, therefore, it is necessary
to optimize through the coin-toss operation $\hat{\cal C}$ for the
QW-encoding.\footnote{It should be noted that the biased coin-toss
operator $\hat{\cal B}$ is dependent also on the fair coin-toss
operator $\hat{\cal C}$.}

\section{\label{sec:concl}Conclusion and Discussions}
In summary, we have shown that by implementing QWs in phase space
for a qumode, it is possible to furnish GKP-type encodings for
quantum error-corrections in CV quantum computing. In addition to
demonstrating an encoding through generic unitary QWs that produces
codewords with ``unconventional" profiles, we show further that an
encoding via {\em dissipative}, nonunitary QWs can generate codeword
states similar to the standard GKP ones. We examine the performance
of the dissipative QW-codewords for error corrections and find that
they do better than the standard GKP codewords by a small amount. In
view of this result, it is promising that with optimized coin-toss
transformations, one may find QW-codewords that perform even better.
Although platforms for implementing QWs may not be easily tailored
to the needs of universal quantum computing \cite{Ma14}, we have
proposed an architecture for bridging this gap through cQED systems.

Throughout this work we have based our encoding scheme on the
discrete-time quantum walk (DTQW). It is pertinent to enquire
whether one could, instead, resort to continuous-time quantum walk
(CTQW) for the task or not. Despite the fact that it is possible to
map between DTQW and CTQW in 1D space \cite{St06}, due to the lack
of the coin degree of freedom in CTQW, we find DTQW provides a
framework that connects more directly to the encoding that we wish
to implement. Nevertheless, it remains an interesting open question
as to whether CTQW can offer additional advantages to the
QW-encoding or not.

Although we have focused primarily on engineering the GKP codeword
states, our work indeed uncovers a new avenue to accessing general
GKP-type ``grid states". For instance, as pointed out earlier,
extending our encoding scheme to the case of {\em two-dimensional}
QWs can be useful for rectifying the undesirable momentum
distributions of the codewords in Fig.~\ref{fig:generic}. At the
same time, this may also help enhance the versatility of the
QW-encoding scheme. In view of the multitude of walk patterns
available for QWs \cite{Ke03,Ma14}, the QW-encoding scheme proposed
in this work thus opens up a new dimension for the GKP encoding that
is yet to be explored. In particular, it may offer a unified
framework for experimentally generating grid states for quantum
error correcting codes in the CV regime.

\begin{acknowledgements}
We are very grateful for Prof. Stephen Barnett's kind help and
insightful suggestions. We also thank Profs. Tzu-Chieh Wei and
Dian-Jiun Han for valuable discussions. This research is supported
by the Ministry of Science and Technology of Taiwan through grants
MOST 107-2112-M-194-002 and MOST 108-2627-E-008-001.
\end{acknowledgements}

\begin{appendix}
\section{\label{sec:uv} Formulas for the amplitudes $u_N^{(\epsilon)}(n)$ and
$v_N^{(\epsilon)}(n)$ in the codeword states}
We provide here explicit expressions for the amplitudes
$u_N^{(\epsilon)}(n)$ and $v_N^{(\epsilon)}(n)$ in \eref{psi_N_form}
for constructing the generic and the dissipative QW codeword states.
\subsection{\label{sec:uv1}Generic QW codeword states}
In the generic (unitary) case if the initial state is a
squeezed-vacuum state $|0\rangle_r$ along with the coin configuration $|R\rangle$,
after $N$ steps of QW one has for $n \neq\pm N$ \cite{Am01}
\begin{eqnarray}
u_N^{(R)}(n)
&=& \frac{1}{\sqrt{2^N}} \sum_{k=0}^{k_u}
\left(
  \begin{array}{c}
     \frac{N-n-2}{2}\\
     k\\
  \end{array}
\right)
\left(
  \begin{array}{c}
    \frac{N+n}{2}\\
    k+1\\
  \end{array}
\right)
(-1)^{\frac{N-n}{2}-k-1} \, ,
\nonumber\\
v_N^{(R)}(n)
&=& \frac{1}{\sqrt{2^N}} \sum_{k=0}^{k_v}
\left(
  \begin{array}{c}
     \frac{N-n-2}{2}\\
     k\\
  \end{array}
\right)
\left(
  \begin{array}{c}
    \frac{N+n}{2}\\
    k\\
  \end{array}
\right)
(-1)^{\frac{N-n}{2}-k-1} \, ,
\label{uv_H1}
\end{eqnarray}
where the upper bounds for the summations are
\begin{eqnarray}
k_u \equiv \min\left\{\frac{N-n-2}{2},\frac{N+n-2}{2}\right\}
\quad{\rm and}\quad
k_v \equiv \min\left\{\frac{N-n-2}{2},\frac{N+n}{2}\right\}\,.
\label{kHV1}
\end{eqnarray}
For the boundary points $n=\pm N$, one finds
\begin{eqnarray}
u_N^{(R)}(+N) &=& \frac{1}{\sqrt{2^N}}\, , \quad u_N^{(R)}(-N) = 0\, , \quad
\nonumber \\
v_N^{(R)}(+N) &=& 0\,, \quad v_N^{(R)}(-N) = \frac{(-1)^{N-1}}{\sqrt{2^N}} \, .
\label{uv_N1}
\end{eqnarray}

In the case of a squeezed-vacuum state with coin configuration $|L\rangle$
initially, one finds after $N$ steps of QW for $n \neq\pm N$
\begin{eqnarray}
u_N^{(L)}(n)
&=& \frac{1}{\sqrt{2^N}} \sum_{k=0}^{k_u'}
\left(
  \begin{array}{c}
     \frac{N+n-2}{2}\\
     k\\
  \end{array}
\right)
\left(
  \begin{array}{c}
    \frac{N-n}{2}\\
    k\\
  \end{array}
\right)
(-1)^{\frac{N-n}{2}-k} \, ,
\nonumber\\
v_N^{(L)}(n)
&=& \frac{1}{\sqrt{2^N}} \sum_{k=0}^{k_v'}
\left(
  \begin{array}{c}
     \frac{N+n-2}{2}\\
     k\\
  \end{array}
\right)
\left(
  \begin{array}{c}
    \frac{N-n}{2}\\
    k+1\\
  \end{array}
\right)
(-1)^{\frac{N-n}{2}-k-1} \, ,
\label{uv_H2}
\end{eqnarray}
where the upper bounds for the summations are
\begin{eqnarray}
k_u' \equiv \min\left\{\frac{N+n-2}{2},\frac{N-n}{2}\right\}
\quad{\rm and}\quad
k_v' \equiv \min\left\{\frac{N+n-2}{2},\frac{N-n-2}{2}\right\}\,.
\label{kHV2}
\end{eqnarray}
For the boundary points $n=\pm N$, one gets
\begin{eqnarray}
u_N^{(L)}(+N) &=& \frac{1}{\sqrt{2^N}}\, , \quad u_N^{(L)}(-N) = 0\, , \quad
\nonumber \\
v_N^{(L)}(+N) &=& 0\,, \quad v_N^{(L)}(-N) = \frac{(-1)^{N}}{\sqrt{2^N}} \, .
\label{uv_N2}
\end{eqnarray}

\subsection{\label{sec:uv2}Dissipative QW codeword states}
In the dissipative case, as pointed out in the text, the $R$ and the
$L$-components are now symmetric. Therefore, for both types of
initial states $|0\rangle_r|R\rangle$ and $|0\rangle_r|L\rangle$,
one has the same result for the amplitudes after $N$ steps of QW. We
can thus drop the superscripts $\epsilon$ for coin configurations in
the amplitudes $u_N^{(\epsilon)}$ and $v_N^{(\epsilon)}$ here. For
$n \neq\pm N$ we find
\begin{eqnarray}
u_N(n) = \frac{1}{2^N}
\left(
  \begin{array}{c}
     N-1 \\
     \frac{N+n-2}{2}\\
  \end{array}
\right)
\quad \mbox{and} \quad
v_N(n) = \frac{1}{2^N}
\left(
  \begin{array}{c}
     N-1 \\
     \frac{N+n}{2}\\
  \end{array}
\right) \, .
\label{uv_dQW1}
\end{eqnarray}
In the case of the boundary points $n=\pm N$, we get
\begin{eqnarray}
u_N(+N) &=& \frac{1}{2^N}\, , \quad u_N(-N) = 0\, , \quad
\nonumber \\
v_N(+N) &=& 0\,, \quad v_N(-N) = \frac{1}{2^N} \, .
\label{uv_dQW2}
\end{eqnarray}
For the amplitudes $w_N(n)$ in \eref{wN1} that incorporates the projection onto the
diagonal coin-state $|D\rangle$, one can then obtain through
\eref{uv_dQW1} and \eref{uv_dQW2} accordingly.

\section{\label{sec:Pnerr} Calculation for $P_{\mbox{\scriptsize no error}}$}
Here we explain briefly how $P_{\mbox{\scriptsize no error}}$ are
calculated for the data plotted in Fig.~\ref{fig:Pnerr}. It has been
shown by Glancy and Knill in Ref.~\cite{Gl06} that when the shift
errors in the codewords are sufficiently bounded, it is possible to
repeatedly recover the corrupted qubits without Pauli errors. To
find the corresponding probabilities, the codewords are projected
onto the basis set $\{|s,t\rangle\}$ that consist of the ideal GKP
codeword $|0\rangle_L$ in \eref{ideal_GKP} shifted in both $x$ and
$p$ \cite{Gl06}
\begin{eqnarray}
|s,t\rangle \equiv \pi^{-\frac{1}{4}}
\sum_{m=-\infty}^\infty e^{-2imt\sqrt{\pi}} |2m\sqrt{\pi}+s\rangle_x \, ,
\label{st_basis}
\end{eqnarray}
where $-\sqrt{\pi}\le s \le \sqrt{\pi}$ and $-\sqrt{\pi}/2\le t \le
\sqrt{\pi}/2$ due to the periodicity of the $|0\rangle_L$ state. The
probability density for a state $|\psi\rangle$ to have shifts $s$
and $t$ relative to the ideal GKP codeword $|0\rangle_L$ is then
given by $|\langle s,t|\psi\rangle|^2$. As demonstrated in
Ref.~\cite{Gl06}, when the shift errors of a codeword are bounded
within the range $[-\sqrt{\pi}/6,+\sqrt{\pi}/6]$ for both $x$ and
$p$ quadratures, it is then ensured that repeated error corrections
will be successful. Namely, for the dQW-codeword $|0\rangle_{\rm
dQW}$, the probability for repeated error corrections without
incurring errors is
\begin{eqnarray}
P_{\mbox{\scriptsize no error}} \equiv
\int_{-\frac{\sqrt{\pi}}{6}}^{+\frac{\sqrt{\pi}}{6}} ds
\int_{-\frac{\sqrt{\pi}}{6}}^{+\frac{\sqrt{\pi}}{6}} dt
\,\,\left|\langle s,t|0\rangle_{\rm dQW}\right|^2 \,.
\label{Pnerr_int}
\end{eqnarray}
Following Eqs.~\eref{psi_N_dQW1}--\eref{psi_N_dQW} and
\eref{st_basis}, one finds upon invoking \eref{n_wfs} for
$x_d=\sqrt{\pi}$
\begin{eqnarray}
\langle s,t|0\rangle_{\rm dQW}
= \left(\frac{e^{+r}}{\pi Z_N}\right)^{\frac{1}{2}}
\sum_{m=-\infty}^\infty \sideset{}{'}\sum_{n=-N}^N w_N(n)\,
e^{+2imt\sqrt{\pi}} \exp\left[-\frac{s+(2m-n)\sqrt{\pi}}{2e^{-2r}}\right] \, .
\label{prob_den}
\end{eqnarray}
Substituting \eref{prob_den} into \eref{Pnerr_int}, one can obtain
the results displayed in Fig.~\ref{fig:Pnerr} accordingly.

\section{\label{sec:statprep}
Circuit for preparing the delayed state $|\tilde{\psi}_{\rm
in}\rangle$}
Here we explain how the circuit shown in Fig.~\ref{fig:QWcir}(b) for
generating the delayed state \eref{psi_dely} can be derived. To
prepare the delayed state, we first displace the input state
$|\psi_{\rm in}\rangle$ of \eref{psi_0} conditionally in the
following way
\begin{eqnarray}
\left[\hat{D}(-\xi_d) \otimes |R\rangle\langle R|
+ \hat{I} \otimes |L\rangle\langle L|\right] |\psi_{\rm in}\rangle
= \alpha\,|-1\rangle_r|R\rangle + \beta\,|0\rangle_r|L\rangle \, .
\label{cDispl}
\end{eqnarray}
Let us next consider a ``biased" coin-toss operation $\hat{\cal B}$,
which induces the map
\begin{eqnarray}
|R\rangle \stackrel{\hat{\cal B}}{\longrightarrow} |R\rangle
\quad\mbox{and}\quad
|L\rangle \stackrel{\hat{\cal B}}{\longrightarrow} \hat{\cal C}\,|L\rangle
\label{B-coin}
\end{eqnarray}
with $\hat{\cal C}$ the original coin-toss operator for the QW.
Suppose one single step of QW is executed for the state
\eref{cDispl} with the biased coin-operator \eref{B-coin} in place
of the original $\hat{\cal C}$. Namely, here we have the ``biased"
walk operator
\begin{eqnarray}
\hat{W}_B \equiv \left[ \hat{D}(+\xi_d) \otimes |R\rangle\langle R| + \hat{D}(-\xi_d)
\otimes |L\rangle\langle L| \right]\, \left(\hat{I} \otimes \hat{\cal B}\right) \, .
\label{W-B}
\end{eqnarray}
It is easy to check that the resulting state would then be the
delayed state $|\tilde{\psi}_{\rm in}\rangle$ of \eref{psi_dely}.
Effectively, here the combined action of the controlled-displacement
\eref{cDispl} and the biased walk-operator \eref{W-B} can be reduced
as follows
\begin{eqnarray}
&& \left[ \hat{D}(+\xi_d) \otimes |R\rangle\langle R| + \hat{D}(-\xi_d)
\otimes |L\rangle\langle L| \right]\, \left(\hat{I} \otimes \hat{\cal B}\right)
\left[\hat{D}(-\xi_d) \otimes |R\rangle\langle R|
+ \hat{I} \otimes |L\rangle\langle L|\right]
\nonumber \\
&=& \left[ \hat{I} \otimes |R\rangle\langle R| + \hat{D}(-2\xi_d)
\otimes |L\rangle\langle L| \right]\, \left(\hat{I} \otimes \hat{\cal B}\right)
\left[ \hat{I} \otimes |R\rangle\langle R|
+ \hat{D}(+\xi_d) \otimes |L\rangle\langle L|\right] \, ,
\label{gate_seq}
\end{eqnarray}
which has the circuit representation of Fig.~\ref{fig:QWcir}(b).


\section{\label{sec:cqed_appdx} Circuit-QED implementations for gate operations in QW-encoding}
This appendix provides details as to how gate operations for
QW-encoding can be realized experimentally in the cQED architecture.
We will first look at realization for the controlled cavity
displacement, and then for the coin-toss operations, including the
biased coin-toss for state preparation and the fair coin-toss for
each step of the QW (see Fig.~\ref{fig:QWcir}). To be specific, for
the coin-toss operations we will focus on the case of dissipative
QW-encoding in our discussions below.

\subsection{\label{sec:cond_displ}The controlled cavity displacement}
\begin{figure*}
\begin{center}
\includegraphics*[width=130mm]{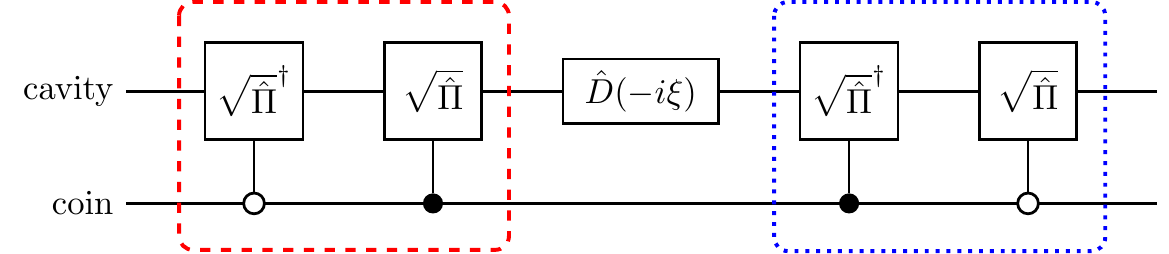}
\caption{Circuit for implementing the controlled cavity displacement
operation, where $\hat{\Pi}\equiv\exp(-i\pi\hat{a}^\dagger\hat{a})$
is the phase-space parity operator for the cavity mode. Here the
controlled gates would enact cavity rotations depending on the coin
qubit being in the state $|0\rangle$ (open circle) or the state
$|1\rangle$ (filled circle). Notice that the gate combination in the
red dashed box is the inverse of that in the blue dotted box.
\label{fig:cond_Displ}}
\end{center}
\end{figure*}

To realize the controlled cavity displacement, here we will adopt a
scheme based on conditional cavity phase shifts as shown in
Fig.~\ref{fig:cond_Displ} \cite{Te16}. As explained in
Sec.~\ref{sec:implm} in the main text, when the cavity-qubit
coupling enters the strong dispersive regime the Jaynes-Cummings
dynamics enables a conditional phase-space rotation of the cavity
mode. For $\chi t=\pi/2$, it follows from \eref{cond_rot} that we
have the operation
\begin{eqnarray}
e^{-i\frac{\pi}{2}\,\hat{a}^\dagger\hat{a}\otimes\hat{Z}} =
\sqrt{\hat{\Pi}} \otimes |0\rangle\langle 0|
+\sqrt{\hat{\Pi}}^\dagger \otimes |1\rangle\langle 1| \, ,
\label{C_pi2}
\end{eqnarray}
where $\hat{\Pi}\equiv\exp(-i\pi\hat{a}^\dagger\hat{a})$ is the
phase-space parity operator for the cavity mode. As in
\eref{cond_rot}, here $|0\rangle$ and $|1\rangle$ stand for,
respectively, the lower and upper qubit levels, which can be
assigned to the coin states, say, as follows
\begin{eqnarray}
|R\rangle\equiv|0\rangle\quad\mbox{and}\quad|L\rangle\equiv|1\rangle\, .
\label{RL_logic}
\end{eqnarray}
It is then easy to verify that the gate sequence in
Fig.~\ref{fig:cond_Displ} would lead to
\begin{eqnarray}
\hat{D}(-\xi) \otimes |0\rangle\langle 0|
+\hat{D}(+\xi) \otimes |1\rangle\langle 1|
= \left[ \hat{I} \otimes |0\rangle\langle 0|
+\hat{D}(+2\xi) \otimes |1\rangle\langle 1| \right]
\hat{D}(-\xi)
\, , \label{C_displ}
\end{eqnarray}
which furnishes the desired controlled cavity displacement
operation. Notice that the pair of gates in the blue dotted box in
Fig.~\ref{fig:cond_Displ} combine to enact the operation
\eref{C_pi2}, while those in the red dashed box yield its inverse.

To realize the gate sequence of Fig.~\ref{fig:cond_Displ}
experimentally, the unconditional cavity displacement
$\hat{D}(-i\xi)$ can be carried out using a short unselective
microwave pulse. For the conditional cavity rotation \eref{C_pi2},
as \eref{cond_rot} suggests, free evolution of the dispersive
Hamiltonian by a waiting time $t=\pi/2\chi$ will furnish the gate
automatically \cite{Le13,Vl13,Le13b}. Similarly, the inverse of
\eref{C_pi2} can be done by free evolution with a waiting time
$t=3\pi/2\chi$, or by conjugating \eref{C_pi2} with Pauli-$X$ gates
for the coin qubit. Alternatively, the controlled cavity
displacement can also be effected using a microwave driving field
resonant with one of the qubit levels \cite{Te16,Vl13,Le13b}. The
resulting qubit-cavity dynamics will then induce cavity
displacements conditioned on the qubit states. Nevertheless, this
approach would require a longer gate operation time.

\subsection{\label{sec:biased_coin}The biased coin-toss operation}
We shall now explain how the biased coin-toss operation for
QW-encoding can be achieved through realizing the corresponding
positive operator-valued measure (POVM). To be specific, we will be
considering the biased coin-toss $\hat{\cal B}_D$ of \eref{B_D} for
the dissipative QW-encoding. To construct a POVM for this operation,
we start by noting that the corresponding measurement operator reads
\cite{Wi10}
\begin{eqnarray}
\hat{M}_D \propto \left(|0\rangle\langle 0| + \frac{|1\rangle\langle +|}{\sqrt{2}}\right) \, ,
\label{M_D}
\end{eqnarray}
where we have denoted the coin states as in \eref{RL_logic}, and
hence
\begin{eqnarray}
|D\rangle\equiv\frac{|0\rangle+|1\rangle}{\sqrt{2}} \equiv|+\rangle \, .
\label{D_logic}
\end{eqnarray}
The effect operator for \eref{M_D} is thus
\begin{eqnarray}
\hat{F}_D=\hat{M}_D^\dagger\hat{M}_D \propto \left(|0\rangle\langle 0| + \frac{|+\rangle\langle +|}{2}\right) \, .
\label{F_D}
\end{eqnarray}
With these observations, we are then able to construct a properly
normalized two-element POVM $\{\hat{F}_1,\hat{F}_2\}$ for our
purpose, which has the elements
\begin{eqnarray}
\hat{F}_1 = \sqrt{\frac{2}{3}} \left(|0\rangle\langle 0| + \frac{|+\rangle\langle +|}{2}\right)
\quad\mbox{and}\quad
\hat{F}_2 = \sqrt{\frac{2}{3}} \left(|1\rangle\langle 1| + \frac{|-\rangle\langle -|}{2}\right) \, ,
\label{povm}
\end{eqnarray}
where $|-\rangle\equiv(|0\rangle-|1\rangle)/\sqrt{2}$, as usual. In
realizations for the POVM \eref{povm}, detection of $\hat{F}_1$ thus
signals successful implementation of the biased coin-toss
transformation \eref{B_D}.

According to Neumark's theorem \cite{Pe93}, for any POVM in a state
space, one can always realize it through orthogonal projective
measurements in an extended state space. For the non-projective POVM
\eref{povm} here, a systematic construction for such ``Neumark
extension" has been developed for photonic qubits \cite{AP05}, which
has recently been extended to circuit architectures \cite{YB19}. To
carry out the scheme, one starts by finding the singular-value
decompositions (SVD's) for the measurement operators $\{\hat{M}_1,
\hat{M_2}\}$ for the POVM \eref{povm}, namely,
\begin{eqnarray}
\hat{M}_1 &=& \sqrt{\frac{2}{3}} \left(|0\rangle\langle 0| +
\frac{1}{\sqrt{2}}|1\rangle\langle +|\right)
= \sqrt{\frac{2}{3}} \left(
                \begin{array}{cc}
                  1 & 0 \\
                  \frac{1}{2} & \frac{1}{2} \\
                \end{array}
              \right) \,,
\nonumber\\
\hat{M}_2 &=& \sqrt{\frac{2}{3}} \left(|1\rangle\langle
1| + \frac{1}{\sqrt{2}}|0\rangle\langle -|\right)
= \sqrt{\frac{2}{3}} \left(
                \begin{array}{cc}
                  \frac{1}{2} & -\frac{1}{2} \\
                  0 & 1 \\
                \end{array}
              \right) \,,
\label{m1m2}
\end{eqnarray}
where the matrices are found with respect to the basis
$\{|0\rangle,|1\rangle\}$. The results for the SVD's are
\begin{eqnarray}
\hat{M}_i = \hat{V} \hat{D}_i \hat{U} \, , \quad \mbox{$i=1,2$} \, ,
\label{M_SVD}
\end{eqnarray}
where
\begin{eqnarray}
\hat{V} &=& \frac{1}{\sqrt{10+2\sqrt{5}}} \left(
                \begin{array}{cc}
                  -2 & 1+\sqrt{5} \\
                  1+\sqrt{5} & 2 \\
                \end{array}
              \right) \,,
\nonumber\\
\hat{U} &=&  \frac{1}{\sqrt{10+4\sqrt{5}}} \left(
                \begin{array}{cc}
                  -1 & 2+\sqrt{5} \\
                  2+\sqrt{5} & 1 \\
                \end{array}
              \right) \,,
\nonumber\\
\hat{D}_1 &=&  \frac{1}{2\sqrt{3}} \left(
                \begin{array}{cc}
                  \sqrt{5}-1 & 0 \\
                  0 & \sqrt{5}+1 \\
                \end{array}
              \right)
              \equiv
              \left(
                \begin{array}{cc}
                  \cos\theta_1 & 0 \\
                  0 & \cos\theta_2 \\
                \end{array}
              \right) \,,
\nonumber\\
\hat{D}_2 &=&  \frac{1}{2\sqrt{3}} \left(
                \begin{array}{cc}
                  \sqrt{5}+1 & 0 \\
                  0 & \sqrt{5}-1 \\
                \end{array}
              \right)
              \equiv
              \left(
                \begin{array}{cc}
                  \sin\theta_1 & 0 \\
                  0 & \sin\theta_2 \\
                \end{array}
              \right) \,.
\label{VDU}
\end{eqnarray}
Notice that although $\theta_2=\pi/2-\theta_1$ here, we keep both
angles explicit in conformity with the notations of
Ref.~\cite{YB19}.

\begin{figure*}
\begin{center}
\includegraphics*[width=130mm]{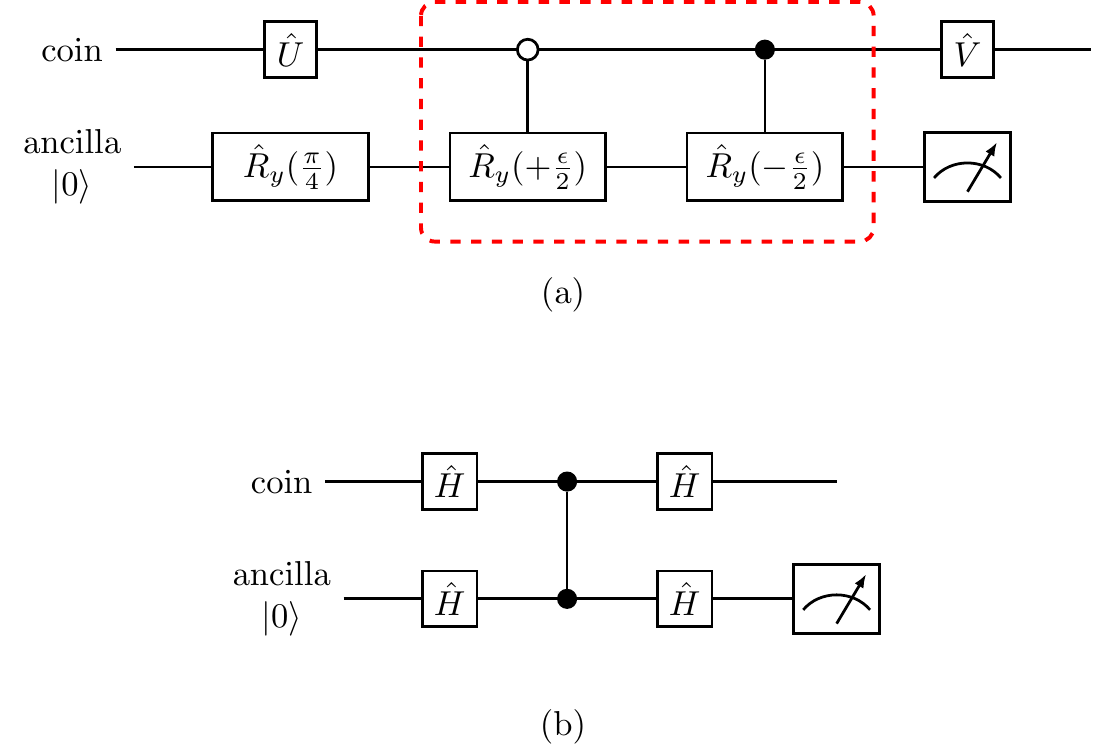}
\caption{Circuits for implementing (a) the biased coin-toss
transformation $\hat{\cal B}_D$ and (b) the fair coin-toss
transformation $\hat{\cal C}_D$ for dissipative QW-encoding. For the
controlled gates in the red dashed box in (a), the gates
$\hat{R}_y(\pm\epsilon/2)$ would act on the ancilla qubit
conditioned on the coin qubit being in the state $|0\rangle$ (open
circle) or the state $|1\rangle$ (filled circle). In both panels,
the measuring device stands for Pauli-$Z$ measurement for the
ancilla qubit. \label{fig:dQW_coins}}
\end{center}
\end{figure*}

Following the scheme of Ref.~\cite{YB19}, one can now construct a
circuit for realizing the POVM \eref{povm} as shown in
Fig.~\ref{fig:dQW_coins}(a), where we have denoted the angle
$\epsilon\equiv(\theta_1-\theta_2)=\cos^{-1}(2/3)\simeq
48.19^{\circ}$ and the operator for single-qubit $y$-rotation by
angle $\phi$ as
\begin{eqnarray}
\hat{R}_y(\phi) \equiv e^{-i\hat{Y}\phi}
\label{Ry}
\end{eqnarray}
with $\hat{Y}$ the Pauli-$Y$ operator. Since both measurement
operators $\hat{M}_i$ share the same unitary operator $\hat{V}$ for
their SVD's in \eref{M_SVD}, here we are able to simplify our
circuit in comparison with that of Ref.~\cite{YB19}. As one can
verify, for an arbitrary coin state $|\psi_c\rangle$ and initial
ancilla state $|0\rangle$, the circuit of
Fig.~\ref{fig:dQW_coins}(a) induces the following map prior to the
ancilla measurement
\begin{eqnarray}
|\psi_c\rangle|0\rangle \rightarrow
\hat{M}_1|\psi_c\rangle|0\rangle + \hat{M}_2|\psi_c\rangle|1\rangle \, ,
\label{M_map}
\end{eqnarray}
where $\hat{M}_1$, $\hat{M}_2$ are the measurement operators of
\eref{m1m2}. Therefore, when Pauli-$Z$ measurement on the ancilla
yields outcome $|0\rangle$, it indicates that the biased coin-toss
transformation \eref{B_D} over the coin qubit has been implemented
successfully.

Experimentally, the circuit in Fig.~\ref{fig:dQW_coins}(a) for
implementing the biased coin-toss should be accessible to current
technologies with cQED systems \cite{We17}. A possible setup for
this purpose is illustrated in the auxiliary part of
Fig.~\ref{fig:cqed}, where a cavity bus resonator ($C_2$) is used to
couple the coin (transmon) qubit ($Q_c$) and the ancilla (transmon)
qubit ($Q_a$), and another cavity ($C_3$) is used to read out and
reset the ancilla state. For the single-qubit unitaries in
Fig.~\ref{fig:dQW_coins}(a), they can be achieved by driving the
transmon qubits with resonant microwave pulses \cite{Bl04,We17}. For
the two-qubit controlled gates enclosed in the red dashed box, they
are exactly the gates realized in the experiments of
Ref.~\cite{Gr13}.\footnote{However, it should be noted that we are
using a different convention for the qubit rotation operators here,
such as $\hat{R}_y(\phi)$, from that in Ref.~\cite{Gr13}. Here we
are following the convention of Ref.~\cite{YB19} for ease in
comparing with the general formulas for POVM realizations there.} We
note that this gate combination can be expressed as
\begin{eqnarray}
\hat{R}_x\!\!\left(\!-\frac{\pi}{4}\right)\,\hat{C}_Z\!\!\left(\frac{\epsilon}{2}\right)
\,\hat{R}_x\!\!\left(\!+\frac{\pi}{4}\right)\,,
\label{decomp}
\end{eqnarray}
where we have defined the conditional $z$-rotation
\begin{eqnarray}
\hat{C}_Z(\phi) \equiv
|0\rangle\langle 0| \otimes \hat{R}_z(\phi) +
|1\rangle\langle 1| \otimes \hat{R}_z(-\phi) \, .
\label{C_phi}
\end{eqnarray}
As in \eref{Ry}, here we have denoted $\hat{R}_x(\phi) \equiv
e^{-i\hat{X}\phi}$ for $x$-rotation of the ancilla qubit and
$\hat{R}_z(\phi) \equiv e^{-i\hat{Z}\phi}$ for its $z$-rotation,
with $\hat{X}$ and $\hat{Z}$ the corresponding Pauli operators. In
the experiment of \cite{Gr13}, the conditional $z$-rotation is
realized by first transferring the state of the control qubit (in
our case, the coin qubit) to the bus cavity. A photon-number
dependent $z$-rotation over the target qubit (the ancilla qubit for
us) is then enacted through free evolution of the dispersive
Hamiltonian. After a waiting time, the cavity state is then
transferred back to the control qubit, which completes the gate
operation $\hat{C}_Z(\phi)$.

An alternate approach to realizing the conditional $z$-rotation
\eref{C_phi} is to exploit the two-qubit higher excitation levels to
generate tunable dynamical phase differences amongst the
computational states \cite{Di09}. This is achieved by flux biasing
one of the qubits, so that its transition frequency is brought to
the proximity of an anti-level-crossing in the two-qubit spectrum.
The corresponding time evolution thus results in the following
effective evolution operator for the two-qubit computational states
$\{|00\rangle,|01\rangle,|10\rangle,|11\rangle\}$
\cite{We17,Di09}\footnote{This two-qubit operator is referred to as
the conditional phase gate in Ref.~\cite{Di09}.}
\begin{eqnarray}
\hat{U} \equiv |00\rangle\langle 00|  + e^{i\phi_{01}} |01\rangle\langle 01|
+ e^{i\phi_{10}} |10\rangle\langle 10|  + e^{i\phi_{11}} |11\rangle\langle 11|  \, ,
\label{U_phi}
\end{eqnarray}
where $\phi_{ij}$ is the dynamical phase of level $|ij\rangle$, and
$\phi_{11}=(\phi_{01}+\phi_{10}-\phi_\zeta)$ with $\phi_\zeta$
adjustable through flux bias control. In the case with
$\phi_{01}=\phi_{10}=\epsilon$ and $\phi_\zeta=2(\epsilon-\pi)$
(thus $\phi_{11}=2\pi$), we have from \eref{U_phi}
\begin{eqnarray}
\hat{U} \mapsto e^{i\frac{\epsilon}{2}} \hat{C}_Z\!\!\left(\frac{\epsilon}{2}\right) \, ,
\label{U_phi2}
\end{eqnarray}
which is the conditional $z$-rotation we wish to implement in
\eref{decomp} up to a global phase. And if we have
$\phi_{01}=\phi_{10}=2\pi$ and $\phi_\zeta=3\pi$ (thus
$\phi_{11}=\pi$), we then have \cite{Di09}
\begin{eqnarray}
\hat{U} \mapsto |0\rangle\langle 0| \otimes \hat{I} + |1\rangle\langle 1| \otimes \hat{Z} \, ,
\label{U_phi3}
\end{eqnarray}
which furnishes the controlled-$Z$ gate. Therefore, by means of flux
bias control, one can realize different two-qubit gates through the
time-evolution operator \eref{U_phi}. This flexibility makes the
construction an ideal choice for implementing the coin-toss
operations, not only for the biased one, but also for the fair one,
which we shall now discuss.

\subsection{\label{sec:fair_coin}The fair coin-toss operation}
We now turn to the realization for the fair coin-toss $\hat{\cal
C}_D$ of \eref{HD-coin} for dissipative QW-encoding. Denoting the
coin states as in \eref{RL_logic} and \eref{D_logic} above, we see
that the gate $\hat{\cal{C}}_D$ is simply the projector for the
Pauli-$X$ eigenstate $|+\rangle$. The circuit for realizing this
gate is thus well known (see, e.g., Ref.~\cite{NC00}) and is given
in Fig.~\ref{fig:dQW_coins}(b). It is easy to show that when the
ancilla qubit is detected in the state $|0\rangle$, the coin state
would be projected onto the state $|+\rangle$, implying successful
implementation of the gate $\hat{\cal C}_D$.

To realize the circuit here experimentally, as discussed in the
preceding subsection, the controlled-$Z$ gate can be implemented
through flux bias control on one of the qubits. Therefore, along
with microwave pulses enacting the Hadamard gates and Pauli-$Z$
measurement for the ancilla qubit, one would be able to furnish the
desired fair coin-toss operation $\hat{\cal C}_D$ in accordance with
the circuit in Fig.~\ref{fig:dQW_coins}(b).

\end{appendix}


\end{document}